\documentclass[12pt]{article}
\usepackage{a4wide,epsfig}

\usepackage{exscale}
\usepackage{axodraw}
\usepackage{amsmath}
\usepackage{slashed}

\newcommand{\agt}{\rlap{\lower 3.5 pt \hbox{$\mathchar \sim$}} \raise 1pt
 \hbox {$>$}}
\newcommand{\alt}{\rlap{\lower 3.5 pt \hbox{$\mathchar \sim$}} \raise 1pt
 \hbox {$<$}}

\catcode`@=11
\newcount\@tempcntc
\def\@citex[#1]#2{\if@filesw\immediate\write\@auxout{\string\citation{#2}}\fi
  \@tempcnta\z@\@tempcntb\m@ne\def\@citea{}\@cite{\@for\@citeb:=#2\do
    {\@ifundefined
       {b@\@citeb}{\@citeo\@tempcntb\m@ne\@citea\def\@citea{,}{\bf
?}\@warning
       {Citation `\@citeb' on page \thepage \space undefined}}%
    {\setbox\z@\hbox{\global\@tempcntc0\csname b@\@citeb\endcsname\relax}%
     \ifnum\@tempcntc=\z@ \@citeo\@tempcntb\m@ne
       \@citea\def\@citea{,}\hbox{\csname b@\@citeb\endcsname}%
     \else
      \advance\@tempcntb\@ne
      \ifnum\@tempcntb=\@tempcntc
      \else\advance\@tempcntb\m@ne\@citeo
      \@tempcnta\@tempcntc\@tempcntb\@tempcntc\fi\fi}}\@citeo}{#1}}
\def\@citeo{\ifnum\@tempcnta>\@tempcntb\else\@citea\def\@citea{,}%
  \ifnum\@tempcnta=\@tempcntb\the\@tempcnta\else
   {\advance\@tempcnta\@ne\ifnum\@tempcnta=\@tempcntb \else
\def\@citea{--}\fi
    \advance\@tempcnta\m@ne\the\@tempcnta\@citea\the\@tempcntb}\fi\fi}
\catcode`@=12

\begin{document}

\title{
\vskip-3cm{\baselineskip14pt
\centerline{\normalsize DESY 07-003\hfill ISSN 0418-9833}
\centerline{\normalsize hep-ph/0702215\hfill}
\centerline{\normalsize January 2007\hfill}}
\vskip1.5cm
$\mathcal{O}(G_F^2m_t^4)$ two-loop electroweak correction to Higgs-boson decay
to bottom quarks}

\author{Mathias Butensch\"on, Frank Fugel, Bernd A. Kniehl\\
{\normalsize II. Institut f\"ur Theoretische Physik, Universit\"at Hamburg,}\\
{\normalsize Luruper Chaussee 149, 22761 Hamburg, Germany}
}
\date{}
\maketitle

\begin{abstract}
We analytically calculate the dominant two-loop electroweak correction, of
\break
$\mathcal{O}(G_F^2m_t^4)$, to the partial width of the decay of a Higgs boson,
with mass $M_H\ll m_t$, into a bottom-quark pair, and describe the most
important conceptual and technical details of our calculation.
As a by-product of our analysis, we also recover the
$\mathcal{O}(\alpha_sG_Fm_t^{2})$ correction.
Relative to the Born result, the $\mathcal{O}(G_F^2m_t^4)$ correction turns
out to be approximately $+0.047\%$ and, thus, more than compensates the
$\mathcal{O}(\alpha_sG_Fm_t^2)$ one, which amounts to
approximately $-0.022\%$.
\medskip

\end{abstract}

\newpage

\section{Introduction}

The standard model (SM) of elementary particle physics predicts the existence
of a last undiscovered particle, the Higgs boson, whose mass $M_H$ is a free
parameter of the theory.
The direct search for the Higgs boson at the CERN Large Electron-Positron
Collider LEP~2 only led to a lower bound of $M_H>114$~GeV at 95\% confidence
level \cite{Barate:2003sz}.
On the other hand, high-precision measurements, especially at LEP and the SLAC
Linear Collider SLC, were sensitive to the Higgs-boson mass via electroweak
radiative corrections.
These indirect measurements yielded the value
$M_H=\left(85^{+39}_{-28}\right)$~GeV and
an upper limit of $M_H<166$~GeV at 95\% confidence level \cite{LEPEWWG}. 
The vacuum-stability and triviality bounds suggest that
$130\alt M_H\alt 180$~GeV if the SM is valid up to the grand-unification scale
(for a review, see Ref.~\cite{Kniehl:2001jy}).
For these reasons, one hopes to discover the Higgs boson at the CERN Large
Hadron Collider (LHC), which will be capable of producing particles with
masses up to 1~TeV.
The first question after discovering a new scalar particle will be if it
actually is the Higgs boson of the SM, or possibly some particle of an
extended Higgs sector.
Therefore, it is necessary to know the SM predictions for the production and
decay rates of the SM Higgs boson with high precision.
Its decay into a bottom-quark pair is of special interest, as it is by far the
dominant decay channel for $M_H\alt140$~GeV (see, for instance,
Ref.~\cite{Kniehl:1993ay}).

At this point, we wish to summarise the current status of the calculations
of radiative corrections to the $H\to b\overline{b}$ decay width in the
so-called intermediate mass range, defined by $M_W\le M_H\le 2M_W$.
The correction of order ${\cal O}(\alpha_s)$ was first calculated in
Ref.~\cite{Braaten:1980yq}.
The complete one-loop electroweak correction was found in
Ref.~\cite{Kniehl:1991}.
As for the ${\cal O}(\alpha_s^2)$ correction, the leading
\cite{Gorishnii:1991zr} and next-to-leading \cite{Surguladze:1994gc} terms of
the expansion in $m_b^2/M_H^2$ of the diagrams without top quarks are known.
The diagrams containing a top quark can be divided into two classes.
The diagrams containing gluon self-energy insertions were calculated
exactly \cite{Kniehl:1994vq}, while for the double-triangle contributions the
four leading terms of the expansion in $M_H^2/m_t^2$ are known
\cite{Chetyrkin:1995pd}.
In Ref.~\cite{Chetyrkin:1996sr}, the ${\cal O}(\alpha_s^3)$ correction
without top-quark contributions was calculated in the massless limit.
The correction induced by th top quark was subsequently found in
Ref.~\cite{Chetyrkin:1997vj} using an appropriate effective field theory.
As for the correction of order ${\cal O}(\alpha_s G_F m_t^2)$, the universal
part, which appears for any Higgs-boson decay to a fermion pair, was
calculated in Ref.~\cite{Kniehl:1994ph} and the non-universal one, using a
low-energy theorem, in Ref.~\cite{KniehlSpira}.
The latter result was independently found in Ref.~\cite{Kwiatkowski:1994cu}.
Apart from the Higgs-boson decay into a $t\overline{t}$ pair, only the one
into a $b\overline{b}$ pair has such non-universal top-quark-induced
contributions, as bottom is the weak-isospin partner of top.
The universal and non-universal corrections of order
${\cal O}(\alpha_s^2 G_F m_t^2)$ were calculated in
Refs.~\cite{delu} and \cite{Chetyrkin:1996ke}, respectively.
Finally, also a result for the universal correction of order
${\cal O}(G_F^2 m_t^4)$ was published \cite{Djouadi}.

In this paper, we calculate the complete correction of order
${\cal O}(G_F^2 m_t^4)$, including both the universal and non-universal
contributions.
To this end, we formally assume that $M_H\ll m_t$.
This includes the intermediate mass range of the Higgs boson.
Our result for the universal contribution in the on-mass-shell scheme agrees
with the one found in Ref.~\cite{Djouadi}, after correcting an obvious mistake
in the latter paper.
The key results of our calculation were already presented in a brief
communication \cite{prl}.
Here, the full details are exhibited.

Our calculations are performed in 't~Hooft-Feynman gauge.
We adopt the on-mass-shell scheme and regularise the ultraviolet divergences
by means of dimensional regularisation, with $D=4-2\epsilon$ space-time
dimensions and 't~Hooft mass scale $\mu$.
We use the anti-commuting definition of $\gamma_5$.
As a simplification, we take the Cabibbo-Kobayashi-Maskawa quark mixing matrix
to be unity.
The Feynman diagrams are generated and drawn using the program
\texttt{FeynArts} \cite{Hahn:2000kx} and evaluated using the program
\texttt{MATAD} \cite{MATAD}, which is written in the programming language
\texttt{FORM} \cite{FORM}.

In order to check our calculations, we also rederive the correction of order
${\cal O}(\alpha_s G_F m_t^2)$.
Our result agrees with
Refs.~\cite{Kniehl:1994ph,KniehlSpira,Kwiatkowski:1994cu}.
Since this calculation follows the lines of the one leading to the
${\cal O}(G_F^2 m_t^4)$ correction, being actually simpler, we refrain from
going into details with it.

This paper is organised as follows.
In Section~\ref{CapRenSchema}, we describe in detail the renormalisation
procedure underlying our analysis.
In Section~\ref{CapOurCalc}, we present the details of our diagrammatic
calculations.
In Section~\ref{CapNieder}, we explain how a part of our calculations can be
checked through the application of a low-energy theorem.
In Section~\ref{Numerics}, we evaluate the ${\cal O}(G_F^2 m_t^4)$ corrections
numerically and compare them with the ${\cal O}(\alpha_s G_F m_t^2)$ ones.
We conclude with a summary in Section~\ref{CapZusammenfassung}.

\section{Renormalisation procedure}
\label{CapRenSchema}

For the reader's convenience, we present in this section the details of the
renormalisation procedure which has to be carried out.
We derive general expressions for the mass counterterms and wave-function
renormalisation constants in the on-shell scheme, valid for any number
of loops.
Furthermore, we derive the tadpole renormalisation counterterms and describe
the treatment of the corrections due to external legs.
In our calculations, we do not need to consider electric-charge
renormalisation constants, because, to the orders we consider here,
there are no such contributions.

Before going into details, we would like to mention that the expressions for
the mass and wave-function renormalisation constants to be derived here are
only valid for stable particles.
Instable particles do have complex self-energy amplitudes, so that their
resummed propagators have complex poles.
In that case, the renormalisation conditions are more complicated (see, for
instance, Ref.~\cite{Kniehl:1998fn}).
Since all self-energy amplitudes appearing in the calculations of this paper
are real, we can restrict ourselves to the case of stable particles.

\subsection{Mass and wave-function renormalisation}
\label{KapMassPar}

We write the bare masses in the Lagrangian as sums of the renormalised ones
and the mass counterterms.
In the on-shell scheme, we fix this splitting by the requirement that the
renormalised masses are identical to the poles of the propagators including
all radiative corrections.
Furthermore, the wave-function renormalisation constants are obtained as the
residues of the propagators at their poles.

\subsubsection{Higgs-boson mass and wave-function renormalisation}
\label{sec:higgs}

For the amputated one-particle-irreducible self-energy of the Higgs
boson, we write
\begin{equation}
\begin{minipage}{112pt}
\begin{picture}(112,32)
  \GOval(56,16)(16,16)(0){0.882}
  \Text(56,16)[]{1-PI}
  \DashLine(40,16)(0,16){4}
  \DashLine(72,16)(112,16){4}
  \Text(92,19)[b]{$H$}
  \Text(20,19)[b]{$H$}
  \LongArrow(12,11)(28,11)
  \Text(20,9)[t]{$q$}
\end{picture}
\end{minipage}
=i\Sigma_H(q^2).
\end{equation}
Thus, the dressed propagator, including all radiative corrections, becomes
\begin{eqnarray}
S_H^{-1}(q^2)&=&
\begin{minipage}{48pt}
\begin{picture}(48,32)
  \DashLine(0,16)(48,16){4}
\end{picture}
\end{minipage}
+
\begin{minipage}{80pt}
\begin{picture}(80,32)
  \GOval(40,16)(16,16)(0){0.882}
  \Text(40,16)[]{1-PI}
  \DashLine(24,16)(0,16){4}
  \DashLine(56,16)(80,16){4}
\end{picture}
\end{minipage}
+
\begin{minipage}{112pt}
\begin{picture}(112,32)
  \DashLine(0,16)(16,16){4}
  \GOval(32,16)(16,16)(0){0.882}
  \Text(32,16)[]{1-PI}
  \DashLine(48,16)(64,16){4}
  \GOval(80,16)(16,16)(0){0.882}
  \Text(80,16)[]{1-PI}
  \DashLine(96,16)(112,16){4}
\end{picture}
\end{minipage}
+\ldots \nonumber \\
&=& \frac{i}{q^2-M_{H,0}^2}\sum_{n=0}^\infty
\left(i\Sigma_H(q^2)\frac{i}{q^2-M_{H,0}^2}\right)^n
\nonumber \\ 
&=&\frac{i}{q^2-M_{H,0}^2+\Sigma_H(q^2)}.
\label{HiggsPropSum}
\end{eqnarray}

The on-shell renormalisation condition reads
\begin{equation} \label{HiggsMassenBed}
S_H(M_H^2) \stackrel{!}{=} 0.
\end{equation}
Writing the bare mass of the Higgs boson as the sum of the renormalised mass
and a counterterm, $M_{H,0}^2 = M_H^2+\delta M_H^2$, we have
\begin{equation} \label{AusdrDmHq1loop}
\delta M_H^2 = \Sigma_H(M_H^2).
\end{equation}
Here and in the following, it is understood that, in the expression for a
counter\-term, all bare quantities have to be replaced by the renormalised ones
plus the respective counter\-terms.
In the case of the Higgs-boson mass counterterm, this means that
$\Sigma_H(M_H^2)$ has to be expressed in terms of renormalised quantities.
For higher-order expressions, this has to be done iteratively.

Expanding Eq.~(\ref{HiggsPropSum}) about $q^2=M_H^2$ and taking the limit
$q^2\to M_H^2$,
\begin{eqnarray}
S_H^{-1}(q^2)&=&\frac{i}{q^2-M_H^2}\,
\frac{1}{1+\Sigma_H^\prime\left(M_H^2\right)+{\cal O}\left(q^2-M_H^2\right)}
\nonumber\\
&&{}\xrightarrow{q^2\to M_H^2}\frac{iZ_H}{q^2-M_H^2},
\end{eqnarray}
we read off the Higgs-boson wave-function renormalisation constant as
\begin{equation}
Z_H = \frac{1}{1+\Sigma_H^\prime(M_H^2)}.
\label{zh}
\end{equation}
Writing $Z_H= 1+\delta Z_H$ and performing a loop expansion of Eq.~(\ref{zh}),
we have
\begin{eqnarray}
\delta Z_H^{(1)} &=& -\Sigma_H^{(1)\prime}(M_H^2),
\label{AusdrDZH1loop}\\
\delta Z_H^{(2)} &=&
-\Sigma_H^{(2)\prime}(M_H^2) +
\left(\Sigma_H^{(1)\prime}(M_H^2)\right)^2.
\label{AusdrDZH2loop}
\end{eqnarray}
Here and in the following, numbers placed in parentheses as superscripts
specify the loop order of the perturbative expression.

\subsubsection{Fermion mass and wave-function renormalisation}
\label{KapFermMassRen}

The amputated one-particle-irreducible self-energy of fermion $f$ has the form
\begin{equation} 
\begin{minipage}{112pt}
\begin{picture}(112,32)
  \GOval(56,16)(16,16)(0){0.882}
  \Text(56,16)[]{1-PI}
  \ArrowLine(0,16)(40,16)
  \ArrowLine(72,16)(112,16)
  \Text(92,20)[b]{$f$}
  \Text(20,20)[b]{$f$}
  \LongArrow(12,11)(28,11)
  \Text(20,9)[t]{$q$}
\end{picture}
\end{minipage}
=i\Sigma_f(q)=i\slashed{q}\omega_- \Sigma_{f,L}(q^2)
+i\slashed{q}\omega_+ \Sigma_{f,R}(q^2) 
+ im_{f,0} \Sigma_{f,S}(q^2),
\label{FermSelbstDef}
\end{equation}
where $m_{f,0}$ is the bare mass of fermion $f$ and
$\omega_\pm =(1\pm\gamma_5)/2$ are the projectors onto the helicity
eigenstates.

The fermion field $f$ is composed of left- and right-handed components, $l$
and $r$, respectively, as
\begin{equation}
f = l + r , \qquad l = \omega_- f , \qquad r = \omega_+ f.
\end{equation}
In the electroweak theory, $l$ and $r$ interact differently, which has to be
accounted for in the renormalisation procedure.
In terms of these components, the purely fermionic part of the SM Lagrangian
reads:
\begin{equation}
{\cal L} = \overline{f}(i\slashed{\partial}-m_{f,0})f
= i\overline{l}\slashed{\partial}l 
+ i\overline{r}\slashed{\partial}r 
- m_{f,0}\overline{r}l - m_{f,0}\overline{l}r.
\end{equation}
We see that $l$ and $r$ are massless fermion fields with propagators
\begin{equation} 
\begin{minipage}{48pt}
\begin{picture}(48,32)
  \ArrowLine(0,16)(48,16)
  \Text(24,20)[b]{$l$}
  \LongArrow(16,11)(32,11)
  \Text(24,9)[t]{$q$}
\end{picture}
\end{minipage}
 = 
\begin{minipage}{48pt}
\begin{picture}(48,32)
  \ArrowLine(0,16)(48,16)
  \Text(24,20)[b]{$r$}
  \LongArrow(16,11)(32,11)
  \Text(24,9)[t]{$q$}
\end{picture}
\end{minipage}
 =  \frac{i}{\slashed{q}}.
\label{FermFirst}
\end{equation}
In addition, we have the following $r$-$l$ transition vertices:
\begin{equation}
\begin{minipage}{48pt}
\begin{picture}(48,32)
  \ArrowLine(0,16)(24,16)
  \Vertex(24,16){2}
  \ArrowLine(24,16)(48,16)
  \Text(12,20)[b]{$l$}
  \Text(36,20)[b]{$r$}
\end{picture}
\end{minipage}
 = 
\begin{minipage}{48pt}
\begin{picture}(48,32)
  \ArrowLine(0,16)(24,16)
  \Vertex(24,16){2}
  \ArrowLine(24,16)(48,16)
  \Text(12,20)[b]{$r$}
  \Text(36,20)[b]{$l$}
\end{picture}
\end{minipage}
 =  -i m_{f,0}.
\label{FermSecond}
\end{equation}
From Eq.~(\ref{FermSelbstDef}), we read off the amputated
one-particle-irreducible self-energies pertaining to the four different
helicity combinations as
\begin{eqnarray}
\begin{minipage}{80pt}
\begin{picture}(80,32)
  \GOval(40,16)(16,16)(0){0.882}
  \Text(40,16)[]{1-PI}
  \ArrowLine(0,16)(24,16)
  \ArrowLine(56,16)(80,16)
  \Text(68,20)[b]{$l$}
  \Text(12,20)[b]{$l$}
  \LongArrow(4,11)(20,11)
  \Text(12,9)[t]{$q$}
\end{picture}
\end{minipage}
&=&i \slashed{q}\Sigma_{f,L}(q^2),
\nonumber\\
\begin{minipage}{80pt}
\begin{picture}(80,32)
  \GOval(40,16)(16,16)(0){0.882}
  \Text(40,16)[]{1-PI}
  \ArrowLine(0,16)(24,16)
  \ArrowLine(56,16)(80,16)
  \Text(68,20)[b]{$r$}
  \Text(12,20)[b]{$r$}
  \LongArrow(4,11)(20,11)
  \Text(12,9)[t]{$q$}
\end{picture}
\end{minipage}
&=&i \slashed{q}\Sigma_{f,R}(q^2),
\nonumber\\
\begin{minipage}{80pt}
\begin{picture}(80,32)
  \GOval(40,16)(16,16)(0){0.882}
  \Text(40,16)[]{1-PI}
  \ArrowLine(0,16)(24,16)
  \ArrowLine(56,16)(80,16)
  \Text(68,20)[b]{$r$}
  \Text(12,20)[b]{$l$}
  \LongArrow(4,11)(20,11)
  \Text(12,9)[t]{$q$}
\end{picture}
\end{minipage}
&=& 
\begin{minipage}{80pt}
\begin{picture}(80,32)
  \GOval(40,16)(16,16)(0){0.882}
  \Text(40,16)[]{1-PI}
  \ArrowLine(0,16)(24,16)
  \ArrowLine(56,16)(80,16)
  \Text(68,20)[b]{$l$}
  \Text(12,20)[b]{$r$}
  \LongArrow(4,11)(20,11)
  \Text(12,9)[t]{$q$}
\end{picture}
\end{minipage}
=i m_{f,0}\Sigma_{f,S}(q^2).
\label{FermLast}
\end{eqnarray}
Note that above expressions do not yet include the tree-level contributions
from Eqs.~(\ref{FermFirst}) and (\ref{FermSecond}).
Equations~(\ref{FermFirst})--(\ref{FermLast}) are the ingredients out of which
we construct the propagators of the left- and right-handed fields including
all radiative corrections.
This is done in close analogy to the case of $\gamma$-$Z$-mixing (see, e.g.,
Ref.~\cite{Hollik:1988ii}).
To this end, we introduce the propagator-type symbols
\begin{eqnarray}
\begin{minipage}{48pt}
\begin{picture}(48,32)
  \ArrowLine(0,16)(48,16)
  \Curve{(20,19)(28,19)}
  \Curve{(20,13)(28,13)}
  \Text(24,22)[b]{$l$}
\end{picture}
\end{minipage}
:&=&
\begin{minipage}{48pt}
\begin{picture}(48,32)
  \ArrowLine(0,16)(48,16)
  \Text(24,20)[b]{$l$}
\end{picture}
\end{minipage}
+
\begin{minipage}{80pt}
\begin{picture}(80,32)
  \GOval(40,16)(16,16)(0){0.882}
  \Text(40,16)[]{1-PI}
  \ArrowLine(0,16)(24,16)
  \ArrowLine(56,16)(80,16)
  \Text(68,20)[b]{$l$}
  \Text(12,20)[b]{$l$}
\end{picture}
\end{minipage}
+
\begin{minipage}{112pt}
\begin{picture}(112,32)
  \ArrowLine(0,16)(16,16)
  \Text(8,20)[b]{$l$}
  \GOval(32,16)(16,16)(0){0.882}
  \Text(32,16)[]{1-PI}
  \ArrowLine(48,16)(64,16)
  \Text(56,20)[b]{$l$}
  \GOval(80,16)(16,16)(0){0.882}
  \Text(80,16)[]{1-PI}
  \ArrowLine(96,16)(112,16)
  \Text(104,20)[b]{$l$}
\end{picture}
\end{minipage}
+\ldots
\nonumber\\
&=& \frac{i}{\slashed{q}} \sum_{n=0}^\infty \left(
i\slashed{q}\Sigma_{f,L}(q^2)\frac{i}{\slashed{q}} \right)^n
= \frac{i}{\slashed{q}\left(1+\Sigma_{f,L}(q^2)\right)},
\nonumber\\
\begin{minipage}{48pt}
\begin{picture}(48,32)
  \ArrowLine(0,16)(48,16)
  \Curve{(20,19)(28,19)}
  \Curve{(20,13)(28,13)}
  \Text(24,22)[b]{$r$}
\end{picture}
\end{minipage}
:&=&
\begin{minipage}{48pt}
\begin{picture}(48,32)
  \ArrowLine(0,16)(48,16)
  \Text(24,20)[b]{$r$}
\end{picture}
\end{minipage}
+
\begin{minipage}{80pt}
\begin{picture}(80,32)
  \GOval(40,16)(16,16)(0){0.882}
  \Text(40,16)[]{1-PI}
  \ArrowLine(0,16)(24,16)
  \ArrowLine(56,16)(80,16)
  \Text(68,20)[b]{$r$}
  \Text(12,20)[b]{$r$}
\end{picture}
\end{minipage}
+
\begin{minipage}{112pt}
\begin{picture}(112,32)
  \ArrowLine(0,16)(16,16)
  \Text(8,20)[b]{$r$}
  \GOval(32,16)(16,16)(0){0.882}
  \Text(32,16)[]{1-PI}
  \ArrowLine(48,16)(64,16)
  \Text(56,20)[b]{$r$}
  \GOval(80,16)(16,16)(0){0.882}
  \Text(80,16)[]{1-PI}
  \ArrowLine(96,16)(112,16)
  \Text(104,20)[b]{$r$}
\end{picture}
\end{minipage}
+\ldots
\nonumber\\
&=& \frac{i}{\slashed{q}} \sum_{n=0}^\infty \left(
i\slashed{q}\Sigma_{f,R}(q^2)\frac{i}{\slashed{q}} \right)^n
= \frac{i}{\slashed{q}\left(1+\Sigma_{f,R}(q^2)\right)},
\end{eqnarray}
and the vertex-type symbols
\begin{eqnarray}
\begin{minipage}{60pt}
\begin{picture}(60,32)
  \GOval(30,16)(6,6)(0){0.882}
  \Vertex(30,16){2}
  \ArrowLine(0,16)(24,16)
  \ArrowLine(36,16)(60,16)
  \Text(48,20)[b]{$r$}
  \Text(12,20)[b]{$l$}
\end{picture}
\end{minipage}
:&=&
\begin{minipage}{48pt}
\begin{picture}(48,32)
  \ArrowLine(0,16)(24,16)
  \Vertex(24,16){2}
  \ArrowLine(24,16)(48,16)
  \Text(12,20)[b]{$l$}
  \Text(36,20)[b]{$r$}
\end{picture}
\end{minipage}
+
\begin{minipage}{80pt}
\begin{picture}(80,32)
  \GOval(40,16)(16,16)(0){0.882}
  \Text(40,16)[]{1-PI}
  \ArrowLine(0,16)(24,16)
  \ArrowLine(56,16)(80,16)
  \Text(68,20)[b]{$r$}
  \Text(12,20)[b]{$l$}
\end{picture}
\end{minipage}
= im_{f,0}\left(\Sigma_{f,S}(q^2)-1 \right),
\nonumber\\
\begin{minipage}{60pt}
\begin{picture}(60,32)
  \GOval(30,16)(6,6)(0){0.882}
  \Vertex(30,16){2}
  \ArrowLine(0,16)(24,16)
  \ArrowLine(36,16)(60,16)
  \Text(48,20)[b]{$l$}
  \Text(12,20)[b]{$r$}
\end{picture}
\end{minipage}
:&=&
\begin{minipage}{48pt}
\begin{picture}(48,32)
  \ArrowLine(0,16)(24,16)
  \Vertex(24,16){2}
  \ArrowLine(24,16)(48,16)
  \Text(12,20)[b]{$r$}
  \Text(36,20)[b]{$l$}
\end{picture}
\end{minipage}
+
\begin{minipage}{80pt}
\begin{picture}(80,32)
  \GOval(40,16)(16,16)(0){0.882}
  \Text(40,16)[]{1-PI}
  \ArrowLine(0,16)(24,16)
  \ArrowLine(56,16)(80,16)
  \Text(68,20)[b]{$l$}
  \Text(12,20)[b]{$r$}
\end{picture}
\end{minipage}
= im_{f,0}\left(\Sigma_{f,S}(q^2)-1 \right).
\end{eqnarray}
Next, we evaluate the dressed propagator of the left-handed fermion field,
including all radiative corrections, as
\begin{eqnarray}
S_{ll}^{-1}(q) &=&
\begin{minipage}{48pt}
\begin{picture}(48,32)
  \ArrowLine(0,16)(48,16)
  \Curve{(20,19)(28,19)}
  \Curve{(20,13)(28,13)}
  \Text(24,22)[b]{$l$}
\end{picture}
\end{minipage}
+
\begin{minipage}{96pt}
\begin{picture}(96,32)
  \ArrowLine(0,16)(24,16)
  \Text(12,22)[b]{$l$}
  \Curve{(8,19)(16,19)}
  \Curve{(8,13)(16,13)}
  \GOval(30,16)(6,6)(0){0.882}
  \Vertex(30,16){2}
  \ArrowLine(36,16)(60,16)
  \Text(48,22)[b]{$r$}
  \Curve{(42,19)(52,19)}
  \Curve{(42,13)(52,13)}
  \GOval(66,16)(6,6)(0){0.882}
  \Vertex(66,16){2}
  \ArrowLine(72,16)(96,16)
  \Text(84,22)[b]{$l$}
  \Curve{(80,19)(88,19)}
  \Curve{(80,13)(88,13)}
\end{picture}
\end{minipage}
+
\begin{minipage}{168pt}
\begin{picture}(168,32)
  \ArrowLine(0,16)(24,16)
  \Text(12,22)[b]{$l$}
  \Curve{(8,19)(16,19)}
  \Curve{(8,13)(16,13)}
  \GOval(30,16)(6,6)(0){0.882}
  \Vertex(30,16){2}
  \ArrowLine(36,16)(60,16)
  \Text(48,22)[b]{$r$}
  \Curve{(42,19)(52,19)}
  \Curve{(42,13)(52,13)}
  \GOval(66,16)(6,6)(0){0.882}
  \Vertex(66,16){2}
  \ArrowLine(72,16)(96,16)
  \Text(84,22)[b]{$l$}
  \Curve{(80,19)(88,19)}
  \Curve{(80,13)(88,13)}
  \GOval(102,16)(6,6)(0){0.882}
  \Vertex(102,16){2}
  \ArrowLine(108,16)(132,16)
  \Text(120,22)[b]{$r$}
  \Curve{(116,19)(124,19)}
  \Curve{(116,13)(124,13)}
  \GOval(138,16)(6,6)(0){0.882}
  \Vertex(138,16){2}
  \ArrowLine(144,16)(168,16)
  \Text(156,22)[b]{$l$}
  \Curve{(152,19)(160,19)}
  \Curve{(152,13)(160,13)}
\end{picture}
\end{minipage}
+\ldots
\nonumber\\
&=&\frac{i}{\slashed{q}\left( 1 + \Sigma_{f,L}(q^2)
  \right)}\sum_{n=0}^\infty \left[ im_{f,0} \left(
\Sigma_{f,S}(q^2)-1\right) \frac{i}{\slashed{q}\left( 1 + \Sigma_{f,R}(q^2)
  \right)}  im_{f,0} \left( \Sigma_{f,S}(q^2)-1\right) 
\right.
\nonumber\\
&&{}\times\left.
\frac{i}{\slashed{q}\left( 1 + \Sigma_{f,L}(q^2) \right)} \right]^n
\nonumber\\
&=&\frac{i\slashed{q}}{1+\Sigma_{f,L}(q^2)}\,\frac{1}{q^2 -
  m_{f,0}^2f(q^2)},
\label{LPropSum}
\end{eqnarray}
where
\begin{equation}
f(q^2)=\frac{(1-\Sigma_{f,S}(q^2))^2}{(1+\Sigma_{f,L}(q^2))
(1+\Sigma_{f,R}(q^2))}.
\end{equation}
In a similar way, we find the dressed propagator of the right-handed fermion
field, including all radiative corrections, to be
\begin{equation} 
S_{rr}^{-1}(q) =
\frac{i\slashed{q}}{1+\Sigma_{f,R}(q^2)}\,\frac{1}{q^2 -
m_{f,0}^2 f(q^2)}.
\label{RPropSum}
\end{equation}
For completeness, we also resum the loop contributions by which a left-handed
field converts into a right-handed one and vice versa.
Proceeding similarly as in Eq.~(\ref{LPropSum}), we obtain
\begin{eqnarray} 
S_{lr}^{-1}(q) &=&
\begin{minipage}{60pt}
\begin{picture}(60,32)
  \ArrowLine(0,16)(24,16)
  \Text(12,22)[b]{$l$}
  \Curve{(8,19)(16,19)}
  \Curve{(8,13)(16,13)}
  \GOval(30,16)(6,6)(0){0.882}
  \Vertex(30,16){2}
  \ArrowLine(36,16)(60,16)
  \Text(48,22)[b]{$r$}
  \Curve{(42,19)(52,19)}
  \Curve{(42,13)(52,13)}
\end{picture}
\end{minipage}
+
\begin{minipage}{132pt}
\begin{picture}(132,32)
  \ArrowLine(0,16)(24,16)
  \Text(12,22)[b]{$l$}
  \Curve{(8,19)(16,19)}
  \Curve{(8,13)(16,13)}
  \GOval(30,16)(6,6)(0){0.882}
  \Vertex(30,16){2}
  \ArrowLine(36,16)(60,16)
  \Text(48,22)[b]{$r$}
  \Curve{(42,19)(52,19)}
  \Curve{(42,13)(52,13)}
  \GOval(66,16)(6,6)(0){0.882}
  \Vertex(66,16){2}
  \ArrowLine(72,16)(96,16)
  \Text(84,22)[b]{$l$}
  \Curve{(80,19)(88,19)}
  \Curve{(80,13)(88,13)}
  \GOval(102,16)(6,6)(0){0.882}
  \Vertex(102,16){2}
  \ArrowLine(108,16)(132,16)
  \Text(120,22)[b]{$r$}
  \Curve{(116,19)(124,19)}
  \Curve{(116,13)(124,13)}
\end{picture}
\end{minipage}
+\ldots
\nonumber\\
&=&\frac{i}{\slashed{q}\left( 1 + \Sigma_{f,L}(q^2) \right)} 
im_{f,0} \left( \Sigma_{f,S}(q^2)-1\right)  \frac{i}{\slashed{q}\left( 1 + 
\Sigma_{f,R}(q^2) \right)}
\nonumber\\
&&{}\times \sum_{n=0}^\infty \left[ im_{f,0} \left( 
\Sigma_{f,S}(q^2)-1\right) \frac{i}{\slashed{q}\left( 1 + 
\Sigma_{f,L}(q^2) \right)}  
im_{f,0} \left( \Sigma_{f,S}(q^2)-1\right)
\right.
\nonumber\\
&&{}\times\left.\frac{i}{\slashed{q}\left( 1 + 
\Sigma_{f,R}(q^2) \right)} \right]^n
\nonumber\\ 
&=& \frac{im_{f,0}(1-\Sigma_{f,S}(q^2))} 
{(1+\Sigma_{f,L}(q^2))(1+\Sigma_{f,R}(q^2))} \,\frac{1}{q^2 - 
m_{f,0}^2f(q^2)}.
\label{PropLtoR}
\end{eqnarray}
Since Eq.~(\ref{PropLtoR}) is symmetric under the interchange of the indices
$L$ and $R$, we also have
\begin{equation}
S_{rl}^{-1}(q)=S_{lr}^{-1}(q).
\label{rightleft}
\end{equation}

We now derive the fermion mass counterterm.
Writing $m_{f,0} = m_f + \delta m_f$, where $m_f$ is the renormalised mass and
$\delta m_f$ is the mass counterterm, and imposing the on-shell
renormalisation condition,
\begin{equation}
\left.S_{ij}(q)u_f(q)\right|_{q^2=m_f^2} \stackrel{!}{=}0,
\label{FermMassA}
\end{equation}
where $ij=ll,rr,lr,rl$ and $u_f(q)$ is the spinor of the incoming fermion $f$,
we obtain
\begin{equation}
\frac{\delta m_f}{m_f} =\frac{1}{\sqrt{f\left(m_f^2\right)}}-1.
\label{FermMassB}
\end{equation}
Expanding Eq.~(\ref{FermMassB}), we find the explicit one- and two-loop
expressions,
\begin{eqnarray}
{\frac{\delta m_f^{(1)}}{m_f}}&=& \frac{1}{2}\Sigma_{f,L}^{(1)}(m_f^2)+
\frac{1}{2}\Sigma_{f,R}^{(1)}(m_f^2)+\Sigma_{f,S}^{(1)}(m_f^2),
\label{DeltaMF1Loop}\\
{\frac{\delta m_f^{(2)}}{m_f}}&=&\frac{1}{2}\Sigma_{f,L}^{(2)}(m_f^2)
+\frac{1}{2}\Sigma_{f,R}^{(2)}(m_f^2)+\Sigma_{f,S}^{(2)}(m_f^2)
-\frac{1}{8}\left(\Sigma_{f,L}^{(1)}(m_f^2)-\Sigma_{f,R}^{(1)}(m_f^2)\right)^2
\nonumber\\
&&{}+\Sigma_{f,S}^{(1)}(m_f^2){\frac{\delta m_f^{(1)}}{m_f}}.
\label{DeltaMF2Loop}
\end{eqnarray}
The one-loop expression of Eq.~(\ref{DeltaMF1Loop}) is well known (see, e.g.,
Ref.~\cite{Kniehl:1991}).
The two-loop expression of Eq.~(\ref{DeltaMF2Loop}) agrees with the one
obtained in Ref.~\cite{Faisst} using an alternative procedure.

Finally, we derive the wave-function renormalisation constants for the
left-handed and right-handed fields.
Expanding Eqs.~(\ref{LPropSum}) and (\ref{RPropSum})--(\ref{rightleft}) about
$\slashed{q}=m_f$ and taking the limit $\slashed{q}\to m_f$, we have
\begin{eqnarray}
S_{ll/rr}^{-1}(q)&=&\frac{i\slashed{q}}{q^2-m_f^2}
\,\frac{1}{\left(1+\Sigma_{f,L/R}(m_f^2)\right)
\left(1-m_f^2\frac{f^\prime(m_f^2)}{f(m_f^2)}\right)
+{\cal O}\left(q^2-m_f^2\right)}
\nonumber\\
&&{}\xrightarrow{q^2\to m_f^2}\frac{i\slashed{q}Z_{f,L/R}}{q^2-m_f^2},
\nonumber\\
S_{lr/rl}^{-1}(q)&=&\frac{im_f}{q^2-m_f^2}
\,\frac{1}{\sqrt{\left(1+\Sigma_{f,L}(m_f^2)\right)
\left(1+\Sigma_{f,R}(m_f^2)\right)}
\left(1-m_f^2\frac{f^\prime(m_f^2)}{f(m_f^2)}\right)
+{\cal O}\left(q^2-m_f^2\right)}
\nonumber\\
&&{}\xrightarrow{q^2\to m_f^2}\frac{im_f\sqrt{Z_{f,L}Z_{f,R}}}{q^2-m_f^2},
\end{eqnarray}
where
\begin{equation}
Z_{f,L/R} = \frac{1}{
\left(1+\Sigma_{f,L/R}(m_f^2)\right) \left( 1 -
m_f^2\frac{f^\prime(m_f^2)}{f(m_f^2)}\right) }.
\label{FermionZ}
\end{equation}
Writing $Z_{f,L/R}= 1+\delta Z_{f,L/R}$ and performing a loop expansion of
Eq.~(\ref{FermionZ}), we have
\begin{eqnarray} 
\delta Z_{f,L}^{(1)} &=& - \Sigma_L^{(1)} 
- \Sigma_L^{(1)\prime} - \Sigma_R^{(1)\prime} - 2\Sigma_S^{(1)\prime},
\label{AusdrDZbl1loop}\\
\delta Z_{f,R}^{(1)} &=& - \Sigma_R^{(1)}
- \Sigma_L^{(1)\prime} - \Sigma_R^{(1)\prime} - 2\Sigma_S^{(1)\prime},
\label{AusdrDZbr1loop}\\
\delta Z_{f,L}^{(2)} &=& - \Sigma_L^{(2)} - \Sigma_L^{(2)\prime} 
- \Sigma_R^{(2)\prime} - 2\Sigma_S^{(2)\prime}
+ \Sigma_L^{(1)}\left(\Sigma_L^{(1)} + 2\Sigma_L^{(1)\prime} 
+ \Sigma_R^{(1)\prime} + 2\Sigma_S^{(1)\prime}\right)
\nonumber\\
&&{}+\Sigma_R^{(1)}\Sigma_R^{(1)\prime}-2\Sigma_S^{(1)}\Sigma_S^{(1)\prime}
+\left(\Sigma_L^{(1)\prime}+\Sigma_R^{(1)\prime}+2\Sigma_S^{(1)\prime}
\right)^2,
\label{AusdrDZbl2loop}\\
\delta Z_{f,R}^{(2)} &=& - \Sigma_R^{(2)} - \Sigma_L^{(2)\prime} 
- \Sigma_R^{(2)\prime} - 2\Sigma_S^{(2)\prime}
+ \Sigma_R^{(1)}\left(\Sigma_R^{(1)} + \Sigma_L^{(1)\prime} 
+ 2\Sigma_R^{(1)\prime} + 2\Sigma_S^{(1)\prime}\right)
\nonumber\\
&&{}+\Sigma_L^{(1)}\Sigma_L^{(1)\prime}-2\Sigma_S^{(1)}\Sigma_S^{(1)\prime}
+\left(\Sigma_L^{(1)\prime}+\Sigma_R^{(1)\prime}+2\Sigma_S^{(1)\prime}
\right)^2.
\label{AusdrDZbr2loop}
\end{eqnarray}
Here, we used the abbreviations
\begin{eqnarray}
\Sigma_X^{(n)}&=&\Sigma_{f,X}^{(n)}(m_f^2),
\nonumber\\
\Sigma_X^{(n)\prime} &=& m_f^2\frac{\partial}{\partial
q^2}\Sigma_{f,X}^{(n)}(q^2) \Big|_{q^2=m_f^2},
\end{eqnarray}
where $X =L,R,S$.
These expressions again agree with Refs.~\cite{Kniehl:1991,Faisst}.

If parity was conserved, we would have $\Sigma_L^f(q^2)=\Sigma_R^f(q^2)$ and
thus recover the structure
\begin{equation}
S_f^{-1}(q) \xrightarrow{q^2\to m_f^2} \frac{iZ_f}{\slashed{q}-m_f},
\end{equation}
which is familiar from quantum electrodynamics.

\boldmath
\subsubsection{$W$-boson mass renormalisation}
\unboldmath

The amputated one-particle-irreducible self-energy of the $W$ boson can be
decomposed into a transverse and a longitudinal part as
\begin{equation}
\begin{minipage}{112pt}
\begin{picture}(112,32)
  \GOval(56,16)(16,16)(0){0.882}
  \Text(56,16)[]{1-PI}
  \Photon(40,16)(0,16){2}{5}
  \Photon(72,16)(112,16){2}{5}
  \Text(92,21)[b]{$W_\nu$}
  \Text(20,20)[b]{$W_\mu$}
  \LongArrow(12,11)(28,11)
  \Text(20,9)[t]{$q$}
\end{picture}
\end{minipage}
= -i\Pi_W^{\mu\nu}(q) 
= -i \left(\Delta^{\mu\nu}\Sigma_{W,T}(q^2) 
+ q^{\mu\nu}\Sigma_{W,L}(q^2)\right),
\label{VecDef}
\end{equation}
where
\begin{eqnarray}
\Delta^{\mu\nu}&=& g^{\mu\nu} - \frac{q^\mu q^\nu}{q^2},
\nonumber\\
q^{\mu\nu}&=& \frac{q^\mu q^\nu}{q^2}.
\end{eqnarray}
Owing to the loop-induced mixing of the $W$ boson with the charged
Higgs-Kibble ghost $\phi$, we must also take into account the
one-particle-irreducible $W\leftrightarrow\phi$ transition amplitudes and the
one-particle-irreducible $\phi$-boson self-energy,
\begin{eqnarray}
\begin{minipage}{112pt}
\begin{picture}(112,32)
  \GOval(56,16)(16,16)(0){0.882}
  \Text(56,16)[]{1-PI}
  \Photon(40,16)(0,16){2}{5}
  \DashLine(72,16)(112,16){4}
  \Text(92,21)[b]{$\phi$}
  \Text(20,20)[b]{$W_\mu$}
  \LongArrow(12,11)(28,11)
  \Text(20,9)[t]{$q$}
\end{picture}
\end{minipage}
&=&iq^{\mu}\Sigma_{W\phi}(q^2),
\nonumber\\
\begin{minipage}{112pt}
\begin{picture}(112,32)
  \GOval(56,16)(16,16)(0){0.882}
  \Text(56,16)[]{1-PI}
  \DashLine(40,16)(0,16){4}
  \Photon(72,16)(112,16){2}{5}
  \Text(92,21)[b]{$W_\mu$}
  \Text(20,20)[b]{$\phi$}
  \LongArrow(12,11)(28,11)
  \Text(20,9)[t]{$q$}
\end{picture}
\end{minipage}
&=&-iq^{\mu}\Sigma_{W\phi}(q^2),
\nonumber\\
\begin{minipage}{112pt}
\begin{picture}(112,32)
  \GOval(56,16)(16,16)(0){0.882}
  \Text(56,16)[]{1-PI}
  \DashLine(40,16)(0,16){4}
  \DashLine(72,16)(112,16){4}
  \Text(92,21)[b]{$\phi$}
  \Text(20,20)[b]{$\phi$}
  \LongArrow(12,11)(28,11)
  \Text(20,9)[t]{$q$}
\end{picture}
\end{minipage}
&=&i\Sigma_\phi(q^2).
\end{eqnarray}
In 't~Hooft-Feynman gauge, the bare propagators of the $W$ and $\phi$ bosons
are given by
\begin{eqnarray}
G_W^{\mu\nu}(q^2) &=& \frac{-ig^{\mu\nu}}{q^2-M_{W,0}^2},
\label{WProp}\\
G_\phi(q^2)&=&\frac{i}{q^2-M_{W,0}^2},
\label{PhiProp}
\end{eqnarray}
with a common bare mass $M_{W,0}$.
In order to obtain the dressed $W$-boson propagator, we proceed in two steps.
In the first step, we resum the one-particle irreducible self-energies of the
$W$ and $\phi$ bosons separately.
In the second step, we systematically combine these results by accommodating
all possible $W\leftrightarrow\phi$ transitions.

The resummation of the one-particle irreducible $W$-boson self-energy leads to
\begin{eqnarray}
\begin{minipage}{48pt}
\begin{picture}(48,32)
  \Photon(0,16)(48,16){2}{5}
  \Curve{(20,20)(28,20)}
  \Curve{(20,12)(28,12)}
  \Text(25,22)[b]{$W$}
\end{picture}
\end{minipage}
:&=&
\begin{minipage}{48pt}
\begin{picture}(48,32)
  \Photon(0,16)(48,16){2}{5}
  \Text(24,20)[b]{$W$}
\end{picture}
\end{minipage}
+
\begin{minipage}{80pt}
\begin{picture}(80,32)
  \GOval(40,16)(16,16)(0){0.882}
  \Text(40,16)[]{1-PI}
  \Photon(0,16)(24,16){2}{3}
  \Photon(56,16)(80,16){2}{3}
  \Text(68,20)[b]{$W$}
  \Text(12,20)[b]{$W$}
\end{picture}
\end{minipage}
+
\begin{minipage}{112pt}
\begin{picture}(112,32)
  \Photon(0,16)(16,16){2}{2}
  \Text(8,20)[b]{$W$}
  \GOval(32,16)(16,16)(0){0.882}
  \Text(32,16)[]{1-PI}
  \Photon(48,16)(64,16){2}{2}
  \Text(56,20)[b]{$W$}
  \GOval(80,16)(16,16)(0){0.882}
  \Text(80,16)[]{1-PI}
  \Photon(96,16)(112,16){2}{2}
  \Text(104,20)[b]{$W$}
\end{picture}
\end{minipage}
+\ldots
\nonumber\\
&=& G_{W,\mu\nu}(q^2) + G_{W,\mu\alpha}(q^2)
\left(-i\Pi_W^{\alpha\beta}(q)\right) G_{W,\beta\nu}(q^2)
\nonumber\\
&&{}+G_{W,\mu\alpha}(q^2)
\left(-i\Pi_W^{\alpha\beta}(q)\right) G_{W,\beta\gamma}(q^2)
\left(-i\Pi_W^{\gamma\delta}(q)\right) G_{W,\delta\nu}(q^2)
+\ldots.
\label{WPropSum} 
\end{eqnarray}
The series in Eq.~(\ref{WPropSum}) may be resummed by inserting
Eqs.~(\ref{VecDef}) and (\ref{WProp}) and exploiting the identities
\begin{eqnarray}
{\Delta^\mu}_\nu{\Delta^\nu}_\rho&=&{\Delta^\mu}_\rho,
\nonumber\\
\Delta^{\mu\nu}q_{\nu\rho}&=& 0,
\nonumber\\
q^{\mu\nu}q_{\nu\rho}&=&{q^\mu}_\rho,
\end{eqnarray}
as follows
\begin{eqnarray}
\begin{minipage}{48pt}
\begin{picture}(48,32)
  \Photon(0,16)(48,16){2}{5}
  \Curve{(20,20)(28,20)}
  \Curve{(20,12)(28,12)}
  \Text(25,22)[b]{$W$}
\end{picture}
\end{minipage}
&=&
G_{W,\mu\alpha}(q^2)
\left[{g^\alpha}_\nu -
\frac{{\Delta^\alpha}_\nu\Sigma_{W,T}(q^2)+{q^\alpha}_\nu\Sigma_{W,L}(q^2)}
{q^2-M_{W,0}^2} 
\right.
\nonumber\\
&&{}+\left.
\frac{{\Delta^\alpha}_\nu\left(\Sigma_{W,T}(q^2)\right)^2
+{q^\alpha}_\nu\left(\Sigma_{W,L}(q^2)\right)^2}
{(q^2-M_{W,0}^2)^2} - \ldots \right]
\nonumber\\
&=& G_{W,\mu\alpha}(q^2)
\left[{g^\alpha}_\nu + {\Delta^\alpha}_\nu \sum_{n=1}^\infty
\left( \frac{-\Sigma_{W,T}(q^2)}{q^2-M_{W,0}^2} \right)^n  +
{q^\alpha}_\nu \sum_{n=1}^\infty \left(
\frac{-\Sigma_{W,L}(q^2)}{q^2-M_{W,0}^2} \right)^n \right]
\nonumber\\
&=&
G_{W,\mu\alpha}(q^2)
\left[ {g^\alpha}_\nu +
{\Delta^\alpha}_\nu\left(\frac{1}{1+\frac{\Sigma_{W,T}(q^2)}{q^2-M_{W,0}^2}}
-1\right)
+{q^\alpha}_\nu\left(\frac{1}{1+\frac{\Sigma_{W,L}(q^2)}{q^2-M_{W,0}^2}}
-1\right)\right]
\nonumber\\
&=&
-i\frac{\Delta_{\mu\nu}}{q^2-M_{W,0}^2+\Sigma_{W,T}(q^2)} 
-i\frac{q_{\mu\nu}}{q^2-M_{W,0}^2+\Sigma_{W,L}(q^2)}
\nonumber\\
&=&\left(S_{W,{\rm pure}}^{-1}\right)_{\mu\nu}(q).
\label{eq:Wpure}
\end{eqnarray}
The resummation of the one-particle-irreducible $\phi$-boson self-energy
proceeds in analogy to the Higgs-boson case discussed in
Section~\ref{sec:higgs} and yields
\begin{eqnarray}
\begin{minipage}{48pt}
\begin{picture}(48,32)
  \DashLine(0,16)(48,16){4}
  \Curve{(20,20)(28,20)}
  \Curve{(20,12)(28,12)}
  \Text(24,22)[b]{$\phi$}
\end{picture}
\end{minipage}
:&=&
\begin{minipage}{48pt}
\begin{picture}(48,32)
  \DashLine(0,16)(48,16){4}
  \Text(24,20)[b]{$\phi$}
\end{picture}
\end{minipage}
+
\begin{minipage}{80pt}
\begin{picture}(80,32)
  \GOval(40,16)(16,16)(0){0.882}
  \Text(40,16)[]{1-PI}
  \DashLine(0,16)(24,16){4}
  \DashLine(56,16)(80,16){4}
  \Text(68,20)[b]{$\phi$}
  \Text(12,20)[b]{$\phi$}
\end{picture}
\end{minipage}
+
\begin{minipage}{112pt}
\begin{picture}(112,32)
  \DashLine(0,16)(16,16){4}
  \Text(8,20)[b]{$\phi$}
  \GOval(32,16)(16,16)(0){0.882}
  \Text(32,16)[]{1-PI}
  \DashLine(48,16)(64,16){4}
  \Text(56,20)[b]{$\phi$}
  \GOval(80,16)(16,16)(0){0.882}
  \Text(80,16)[]{1-PI}
  \DashLine(96,16)(112,16){4}
  \Text(104,20)[b]{$\phi$}
\end{picture}
\end{minipage}
+\ldots
\nonumber\\
&=&\frac{i}{q^2-M_{W,0}^2+\Sigma_\phi(q^2)}.
\label{eq:Phipure}
\end{eqnarray}

The contribution of unmixed $W$-boson propagation in Eq.~(\ref{eq:Wpure})
needs to be complemented by the contribution that emerges by combining it with
the contribution of unmixed $\phi$-boson propagation of Eq.~(\ref{eq:Phipure})
via the one-particle-irreducible $W\leftrightarrow\phi$ transition amplitudes
in all possible ways.
This additional contribution is given by
\begin{eqnarray}
\left(S_{W,{\rm mix}}^{-1}\right)_{\mu\nu}(q)&=&
\begin{minipage}{96pt}
\begin{picture}(96,32)
  \Photon(0,16)(24,16){1.5}{3}
  \Text(13,22)[b]{$W$}
  \Curve{(8,20)(16,20)}
  \Curve{(8,12)(16,12)}
  \GOval(30,16)(6,6)(0){0.882}
  \Text(30,16)[]{\tiny 1PI}
  \DashLine(36,16)(60,16){4}
  \Text(47,22)[b]{$\phi$}
  \Curve{(42,20)(52,20)}
  \Curve{(42,12)(52,12)}
  \GOval(66,16)(6,6)(0){0.882}
  \Text(66,16)[]{\tiny 1PI}
  \Photon(72,16)(96,16){1.5}{3}
  \Text(85,22)[b]{$W$}
  \Curve{(80,20)(88,20)}
  \Curve{(80,12)(88,12)}
\end{picture}
\end{minipage}
+
\begin{minipage}{168pt}
\begin{picture}(168,32)
  \Photon(0,16)(24,16){1.5}{3}
  \Text(13,22)[b]{$W$}
  \Curve{(8,20)(16,20)}
  \Curve{(8,12)(16,12)}
  \GOval(30,16)(6,6)(0){0.882}
  \Text(30,16)[]{\tiny 1PI}
  \DashLine(36,16)(60,16){4}
  \Text(47,22)[b]{$\phi$}
  \Curve{(42,20)(52,20)}
  \Curve{(42,12)(52,12)}
  \GOval(66,16)(6,6)(0){0.882}
  \Text(66,16)[]{\tiny 1PI}
  \Photon(72,16)(96,16){1.5}{3}
  \Text(85,22)[b]{$W$}
  \Curve{(80,20)(88,20)}
  \Curve{(80,12)(88,12)}
  \GOval(102,16)(6,6)(0){0.882}
  \Text(102,16)[]{\tiny 1PI}
  \DashLine(108,16)(132,16){4}
  \Text(120,22)[b]{$\phi$}
  \Curve{(116,20)(124,20)}
  \Curve{(116,12)(124,12)}
  \GOval(138,16)(6,6)(0){0.882}
  \Text(138,16)[]{\tiny 1PI}
  \Photon(144,16)(168,16){1.5}{3}
  \Text(157,22)[b]{$W$}
  \Curve{(152,20)(160,20)}
  \Curve{(152,12)(160,12)}
\end{picture}
\end{minipage}
+\ldots
\nonumber\\
&=&\frac{q_{\mu}\Sigma_{W\phi}(q^2)}{q^2-M_{W,0}^2+\Sigma_{W,L}(q^2)} \,
\frac{i}{q^2-M_{W,0}^2+\Sigma_\phi(q^2)}
\nonumber\\
&&{}\times \sum_{n=0}^\infty\left(\frac{q^2
(\Sigma_{W\phi}(q^2))^2}{\left(q^2-M_{W,0}^2+\Sigma_{W,L}(q^2)\right)
\left(q^2-M_{W,0}^2+\Sigma_\phi(q^2)\right)}\right)^n
\nonumber\\
&&{}\times
\frac{-q_{\nu}\Sigma_{W\phi}(q^2)}{q^2-M_{W,0}^2+\Sigma_{W,L}(q^2)}
\nonumber\\
&=&\frac{-iq_{\mu\nu}}{q^2-M_{W,0}^2+\Sigma_{W,L}(q^2)}\,
 \frac{-q^2}{q^2 - \frac{\left(q^2-M_{W,0}^2+\Sigma_{W,L}(q^2)\right)
\left(q^2-M_{W,0}^2+\Sigma_\phi(q^2)\right)}
{(\Sigma_{W\phi}(q^2))^2}}.
\label{WPropMixCont}
\end{eqnarray}

Adding Eqs.~(\ref{eq:Wpure}) and (\ref{WPropMixCont}), we obtain the fully
dressed  $W$-boson propagator as
\begin{equation}
\left(S_W^{-1}\right)_{\mu\nu}(q)
=\left(S_{W,{\rm pure}}^{-1}\right)_{\mu\nu}(q)
+\left(S_{W,{\rm mix}}^{-1}\right)_{\mu\nu}(q).
\end{equation}
Its inverse is found to be
\begin{eqnarray}
S_W^{\mu\nu}(q)&=&ig^{\mu\nu}(q^2-M_{W,0}^2) +
i\Delta^{\mu\nu}\Sigma_{W,T}(q^2)
+iq^{\mu\nu}\left( \Sigma_{W,L}(q^2)
- \frac{q^2\left(\Sigma_{W\phi}(q^2)\right)^2} 
{q^2-M_{W,0}^2+\Sigma_\phi(q^2)} \right).\qquad
\label{InvVPropSumMix}
\end{eqnarray}
The on-shell renormalisation condition reads
\begin{equation}
\left.S_W^{\mu\nu}(q^2)\epsilon_{W,\nu}(q) \right|_{q^2=M_W^2}
\stackrel{!}{=} 0,
\label{VMassenBed}
\end{equation}
where $\epsilon_W^\mu(q)$ is the polarisation four-vector of an external
$W$ boson.
Writing $M_{W,0}^2=M_W^2+\delta M_W^2$ and exploiting the transversality
property $q^\mu\epsilon_{W,\mu}(q)=0$, we finally have
\begin{equation}
\delta M_W^2 = \Sigma_{W,T}(M_W^2).
\label{AusdrDmwq}
\end{equation}
We note in passing that Eq.~(\ref{AusdrDmwq}) is not influenced by
$W\leftrightarrow\phi$ mixing.

\subsection{External-leg corrections}
\label{CapWFR}

In this section, we discuss the structure of the amputated matrix element
${\cal A}$ for the decay process $H \to b\overline{b}$ and explain how to
obtain from it the transition matrix element ${\cal T}$ by incorporating the
wave-function renormalisation constants.

The general form of ${\cal A}$ reads
\begin{eqnarray}
\lefteqn{
\begin{minipage}{104.5pt}
\begin{picture}(104.5,62)
  \ArrowLine(69.85,39)(104.5,59)
  \ArrowLine(104.5,3)(69.85,23)
  \GOval(56,31)(16,16)(0){0.882}
  \Text(56,31)[]{Amp.}
  \DashLine(40,31)(0,31){5}
  \Text(20,34)[b]{$H$}
  \Text(87.175,16.5)[bl]{$\overline{b}$}
  \Text(87.175,45.5)[tl]{$b$}
  \LongArrow(10,26)(30,26)
  \Text(20,24)[t]{$q_1\!\!+\!q_2$}
  \LongArrow(77.7487,49.3301)(91.6051,57.3301)
  \Text(85,56)[br]{$q_2$}
  \LongArrow(77.7487,12.66987)(91.6051,4.66987)
  \Text(86,7)[tr]{$q_1$}
\end{picture}
\end{minipage}
= i {\cal A}}
\nonumber\\
&=& i \left( {\cal A}_1 + \slashed{q}_1 {\cal A}_2 +
\slashed{q}_2 {\cal A}_3 + \slashed{q}_2\slashed{q}_1 {\cal A}_4 +
\gamma_5 {\cal A}_5 + \gamma_5\slashed{q}_1 {\cal A}_6 +
\gamma_5\slashed{q}_2 {\cal A}_7 + \gamma_5\slashed{q}_2
\slashed{q}_1 {\cal A}_8 \right),
\label{HbbStruktur}
\end{eqnarray}
where $q_1$ and $q_2$ are the four-momenta of the outgoing $\overline{b}$ and
$b$ quarks, respectively, and ${\cal A}_i$ ($i=1,\ldots,8$) are scalar form
factors.
Projecting onto each of these form factors, we observe that, to the orders we
consider in this paper, only two of them are independent.
In fact, we have
\begin{eqnarray}
{\cal A}_2&=&-{\cal A}_3={\cal A}_6=-{\cal A}_7,
\nonumber\\
{\cal A}_4&=&{\cal A}_5={\cal A}_8=0,
\end{eqnarray}
so that ${\cal A}$ collapses to the simple form
\begin{equation} 
{\cal A} = {\cal A}_A + {\cal A}_B \left(
\slashed{q}_2 - \slashed{q}_1 \right)\omega_-,
\label{AAufspaltung}
\end{equation}
where ${\cal A}_A={\cal A}_1$ and ${\cal A}_B=-2{\cal A}_2$.

Then, ${\cal T}$ is obtained by dressing ${\cal A}$ with the renormalised wave
functions of the external legs as
\begin{eqnarray}
{\cal T} &=& \sqrt{Z_H}  \left( \sqrt{Z_{b,R}} \overline{u}_r(q_2,r_2) +
\sqrt{Z_{b,L}} \overline{u}_l(q_2,r_2) \right)  {\cal A}  \left(
\sqrt{Z_{b,R}} v_r(q_1,r_1) + \sqrt{Z_{b,L}} v_l(q_1,r_1) \right)
\nonumber\\
&=& \sqrt{Z_H}  \overline{u}_b(q_2,r_2) \left(\sqrt{Z_{b,R}}\omega_- +
\sqrt{Z_{b,L}}\omega_+\right)  {\cal A}  \left(\sqrt{Z_{b,R}}\omega_+ +
\sqrt{Z_{b,L}}\omega_-\right)  v_b(q_1,r_1),
\label{TAusAundZVor1}
\end{eqnarray}
where $v_b(q_1,r_1)$ and $\overline{u}_b(q_2,r_2)$ denote the spinors of the
outgoing $\overline{b}$ and $b$ quarks with spins $r_1$ and $r_2$,
respectively.
Inserting Eq.~(\ref{AAufspaltung}) into Eq.~(\ref{TAusAundZVor1}), we obtain
the master formula
\begin{equation}
{\cal T} = \sqrt{Z_H}\left(\sqrt{Z_{b,L}Z_{b,R}} {\cal A}_A
+m_bZ_{b,L}{\cal A}_B\right)\overline{u}_b(q_2,r_2)v_b(q_1,r_1).
\label{TAusAundZ}
\end{equation}
Note, that the terms involving $\gamma_5$ vanish upon application of the Dirac
equation.

\subsection{Tadpole renormalisation}
\label{KapTadpoleren}

As is well known (see, for instance, Ref.~\cite{Denner}), one can
introduce a so-called tadpole renormalisation in order to avoid the
calculation of diagrams containing tadpoles.
For the reader's convenience, in this section, we rederive the counterterm
vertices of the tadpole renormalisation along with the counterterm vertices
of the Higgs-boson mass renormalisation.

The tadpole renormalisation concerns only the Higgs part of the SM Lagrangian,
\begin{equation}
{\cal L}_\mathrm{Higgs} = (D_\mu\Phi)^\dagger(D^\mu\Phi)
+\mu^2\Phi^\dagger\Phi-\frac{\lambda}{4}(\Phi^\dagger\Phi)^2,
\label{Lag1}
\end{equation}
where $\Phi$ is a weak-isospin doublet of two complex scalar fields.
The free parameters, $\mu$ and $\lambda$, are chosen in such a way that one
stays with a non-vanishing vacuum expectation value $v$, which is defined by
\begin{equation}
\frac{v^2}{2} = \left| \langle 0|\Phi(x)|0\rangle \right|^2
=\frac{2\mu^2}{\lambda}.
\label{vac}
\end{equation}

If we parameterise
\begin{equation}
\Phi(x) = {\phi^+(x) \choose \frac{1}{\sqrt{2}}\left(v +
H(x)+i\chi(x)\right)}
\end{equation}
and substitute $\mu$ and $\lambda$ by
\begin{eqnarray}
t&=& v\left(\mu^2-\frac{\lambda v^2}{4}\right),
\nonumber\\
M_H^2&=& - \mu^2 + \frac{3\lambda v^2}{4},
\end{eqnarray}
Eq.~(\ref{Lag1}) takes the form
\begin{eqnarray}
{\cal L}_\mathrm{Higgs} &=& \frac{1}{2}(D_\mu H)(D^\mu H) +
\frac{1}{2}(D_\mu\chi)(D^\mu\chi) + (D_\mu\phi^-)(D^\mu\phi^+) 
+ t H -\frac{M_H^2}{2}H^2 
\nonumber\\
&&{}+ \frac{t}{2v}\left(\chi^2+2 \phi^-\phi^+\right)
-\frac{1}{2v}\left(\frac{t}{v}+M_H^2\right)H
\left(H^2+\chi^2+2 \phi^-\phi^+\right)
\nonumber\\
&&{}-\frac{1}{8v^2}\left(\frac{t}{v}+M_H^2\right)
\left(H^2+\chi^2+2 \phi^-\phi^+\right)^2,
\label{Lag2}
\end{eqnarray}
where $\phi^- = \left( \phi^+ \right)^\dagger$.
We see that $M_H$ has the physical meaning of the Higgs-boson mass.
In this step, we did not exploit Eq.~(\ref{vac}), which implies that $t=0$, so
that we could just have emitted all terms containing $t$.
However, as was argued above, it is useful to keep them and to renormalise $t$
along with $M_H^2$ by substituting
\begin{eqnarray}
t&\to& t_0 = 0 + \delta t,
\nonumber\\
M_H^2 &\to& M_{H,0}^2 = M_H^2 + \delta M_H^2
\end{eqnarray}
in Eq.~(\ref{Lag2}).
Notice that Eq.~(\ref{Lag2}) represents a bare Lagrangian, so that $v$, $t$,
and $M_H$ are actually bare parameters.
For consistency, we thus also substitute $v\to v_0$.  
Then, Eq.~(\ref{Lag2}) becomes
\begin{eqnarray}
{\cal L}_\mathrm{Higgs} &=&\frac{1}{2}(D_\mu H)(D^\mu H) +
\frac{1}{2}(D_\mu\chi)(D^\mu\chi) +
(D_\mu\phi^-)(D^\mu\phi^+) -\frac{M_H^2}{2} H^2 
\nonumber\\
&&{}-\frac{M_H^2}{2v_0}H\left(H^2+\chi^2+2\phi^-\phi^+\right)
-\frac{M_H^2}{8v_0^2}\left(H^2+\chi^2+2\phi^-\phi^+\right)^2
\nonumber\\
&&{}+\delta t H-\frac{\delta M_H^2}{2} H^2 
+\frac{\delta t}{2v_0}\left(\chi^2 +2\phi^-\phi^+\right)
-\frac{1}{2v_0}\left(\frac{\delta t}{v_0}+\delta M_H^2\right)H
\nonumber\\
&&{}\times\left(H^2+\chi^2+2\phi^-\phi^+\right)
-\frac{1}{8v_0^2}\left(\frac{\delta t}{v_0}+\delta M_H^2\right)
\left(H^2+\chi^2+2\phi^-\phi^+\right)^2.
\end{eqnarray}
From the terms proportional to $\delta t$ and $\delta M_H^2$, we can read off
the desired counterterm vertices, which we list in Table~\ref{TabCTs}.
\begin{table}
\renewcommand{\arraystretch}{2.25}
\begin{center}
\caption{\label{TabCTs}Counterterm vertices related to the Higgs-boson tadpole
and mass renormalisation.}
\fbox{
\begin{tabular}{l@{}c@{\hspace{1.75cm}}l@{}c}
$H$: & $i \delta t$ & 
$HHHH$: &
$\displaystyle
-i\frac{3}{v_0^2}\left(\frac{\delta t}{v_0} + \delta M_H^2\right)$ \\
$HH$: & $-i \delta M_H^2$ & 
$\chi\chi\chi\chi$: & 
$\displaystyle
-i\frac{3}{v_0^2} \left(\frac{\delta t}{v_0} + \delta M_H^2\right)$ \\
$\chi\chi$: & $\displaystyle i\frac{\delta t}{v_0}$ &
$HH\chi\chi$: &
$\displaystyle 
-i\frac{1}{v_0^2} \left(\frac{\delta t}{v_0} + \delta M_H^2\right)$ \\
$\phi\phi$: & $\displaystyle i\frac{\delta t}{v_0}$ & 
$HH\phi\phi$: &
$\displaystyle
-i\frac{1}{v_0^2} \left(\frac{\delta t}{v_0} + \delta M_H^2\right)$ \\
$HHH$: & $\displaystyle
-i\frac{3}{v_0}\left(\frac{\delta t}{v_0} + \delta M_H^2\right)$ &
$\chi\chi\phi\phi$: & 
$\displaystyle
-i\frac{1}{v_0^2} \left(\frac{\delta t}{v_0} + \delta M_H^2\right)$ \\
$H\chi\chi$: &
$\displaystyle
-i\frac{1}{v_0}\left(\frac{\delta t}{v_0} + \delta M_H^2\right)$ &
$\phi\phi\phi\phi$: &
$\displaystyle
-i\frac{2}{v_0^2}\left(\frac{\delta t}{v_0} + \delta M_H^2\right)$ \\
$H\phi\phi$: &
$\displaystyle
-i\frac{1}{v_0}\left(\frac{\delta t}{v_0} + \delta M_H^2\right)$
\vspace{5pt}
\end{tabular}
}
\end{center}
\end{table}

The Higgs-boson mass renormalisation condition was already discussed in
Section~\ref{sec:higgs}.
As a renormalisation condition for $\delta t$, we set
\begin{equation}
\delta t \stackrel{!}{=} -T,
\label{AusdrDt}
\end{equation}
where $T$ stands for the sum of all amputated one-particle-irreducible
tadpole diagrams,
\begin{equation}
\begin{minipage}{72pt}
\begin{picture}(72,32)
  \GOval(56,16)(16,16)(0){0.882}
  \Text(56,16)[]{1-PI}
  \DashLine(40,16)(0,16){5}
  \Text(20,19)[b]{$H$}
\end{picture}
\end{minipage}
=iT.
\label{DefTadpole}
\end{equation}
As can be seen from Table~\ref{TabCTs}, there is a one-point Higgs-boson
counterterm vertex, $i\delta t$, that forces a cancellation with all
diagrams having a tadpole at its place.
Therefore, upon tadpole renormalisation, one does not have to consider tadpole
diagrams anymore.
However, now one has to take into account all the tadpole counterterm vertices
in Table~\ref{TabCTs}, except for the one mentioned above.

\section{Results}
\label{CapOurCalc}

In this section, we present the details of our actual calculations.
After making some general remarks, we describe in Sections~\ref{SecTree},
\ref{SecOneLoop}, and \ref{CapEW2LoopKorr} the explicit computation of the
decay rate at tree level, at the one-loop order ${\cal O}(G_F m_t^2)$, and at
the two-loop order ${\cal O}(G_F^2 m_t^4)$, respectively.
Section~\ref{SecTree} also contains the expressions for the renormalisation
constants at order ${\cal O}(G_F m_t^2)$, which are needed in the one-loop
and two-loop calculations.

In order to compute the leading large-$m_t$ contributions of the
various two-loop dia\-grams, we apply the asymptotic-expansion technique (for
a careful introduction, see Ref.~\cite{Smirnov}).
However, it turns out that all non-trivial contributions of the self-energy
and $Hb\overline{b}$ vertex diagrams (see Figs.~\ref{DiaW2l},
\ref{DiaHbb2loop}, and \ref{DiaB2l}), which are of leading order in $m_t$,
cancel among themselves or, in case of the $W$-boson self-energy, in
combination with
complete counterterm diagrams
arising form the Higgs-boson tadpole and mass renormalisations.
Specifically, in Fig.~\ref{DiaW2l}, there are non-naive contributions due to
the asymptotic expansion of diagrams (i)--(o) that cancel against diagrams
(p)--(v); in Fig.~\ref{DiaHbb2loop}, the non-naive contributions of diagrams
(a) and (t) cancel; and in Fig.~\ref{DiaB2l} those of the diagrams (e) and
(i) cancel.
After these cancellations, only naive contributions due to diagrams involving
top-quark propagators remain.
Therefore, we can naively expand in all masses and momenta except for the
top-quark mass and retain only the leading terms.
Obviously, this requires the Higgs-boson mass to be smaller than the top-quark
mass, which is compatible with the intermediate-mass range of the Higgs boson,
as mentioned in the Introduction.

The ultraviolet divergences which have to disappear in the final expression
for the decay rate are cancelled through the application of the
renormalisation procedure, which we carry out in the on-mass-shell
renormalisation scheme.
This provides a non-trivial check for our calculations.
As explained in Section~\ref{KapTadpoleren}, we use the counterterm vertices
of Table~\ref{TabCTs} for the Higgs-boson tadpole and mass renormalisations.
However, while we renormalise the Higgs-boson mass already at the Lagrangian
level, we replace all other bare parameters at the end of the calculations
without recourse to any counterterm vertices.
This procedure turns out to be most convenient for our purposes.

As a further check on our calculations, we also rederive the correction of
order\break ${\cal O}(\alpha_s G_F m_t^2)$.
This result is presented in Section~\ref{sec:mixed}.
Finally, we apply a Higgs-boson low-energy theorem \cite{Kniehl:1995tn}, which
allows for an independent calculation of the various $Hb\overline{b}$ diagrams
at order ${\cal O}(G_F^2m_t^4)$.
This is explained in Section~\ref{CapNieder}.

\boldmath
\subsection{Tree-level result and ${\cal O}(G_F m_t^2)$
renormalisation constants}
\label{SecTree}
\unboldmath

\begin{figure}
\begin{center}
\includegraphics[width=0.75\textwidth]{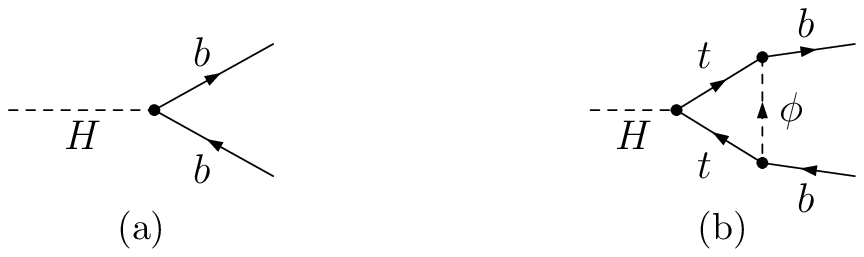}
\end{center}
\caption{\label{DiaHbb}Diagrams contributing to $H\to b\overline{b}$ at (a)
tree level and (b) order ${\cal O}(G_Fm_t^2)$.}
\end{figure}

The tree-level diagram is depicted in Fig.~\ref{DiaHbb}(a).
Using the notation introduced in Eq.~(\ref{AAufspaltung}), the corresponding
amputated matrix element is in bare form written as
\begin{equation}\label{BornUnren}
{\cal A}^{(0)}_{0} = {\cal A}^{(0)}_{A,0} = -\frac{m_{b,0}}{v_0}.
\end{equation}
The tree-level transition matrix element is
\begin{equation}\label{ErgT0}
{\cal T}^{(0)}={\cal A}_0^{(0)}\overline{u}_b(q_2,r_2)v_b(q_1,r_1),
\end{equation}
and the decay rate is
\begin{equation}\label{BornTrans}
\Gamma^{(0)}=\frac{\sqrt{2}N_cG_FM_Hm_b^2}{8\pi}
\left(1-\frac{4m_b^2}{M_H^2}\right)^{3/2},
\end{equation}
where $N_c=3$ is the number of quark colours.
Furthermore, we have introduced Fermi's constant $G_F$ via the Born relation
\begin{equation} \label{Defv}
\frac{1}{v} = 2^{1/4}G_F^{1/2}.
\end{equation}

In the following, we have to renormalise the vacuum expectation value.
Through the order of our calculations, this can be achieved by writing
\cite{Consoli:1989fg}
\begin{equation}
\frac{1}{v_0} = 2^{1/4}G_{F,0}^{1/2},
\end{equation}
with
\begin{equation}
G_{F,0} = G_F \frac{M_W^2}{M_{W,0}^2}.
\end{equation}
Thus, the renormalisation of the vacuum expectation value is reduced to the
one of the $W$-boson mass.

\begin{figure}
\begin{center}
\includegraphics[width=\textwidth]{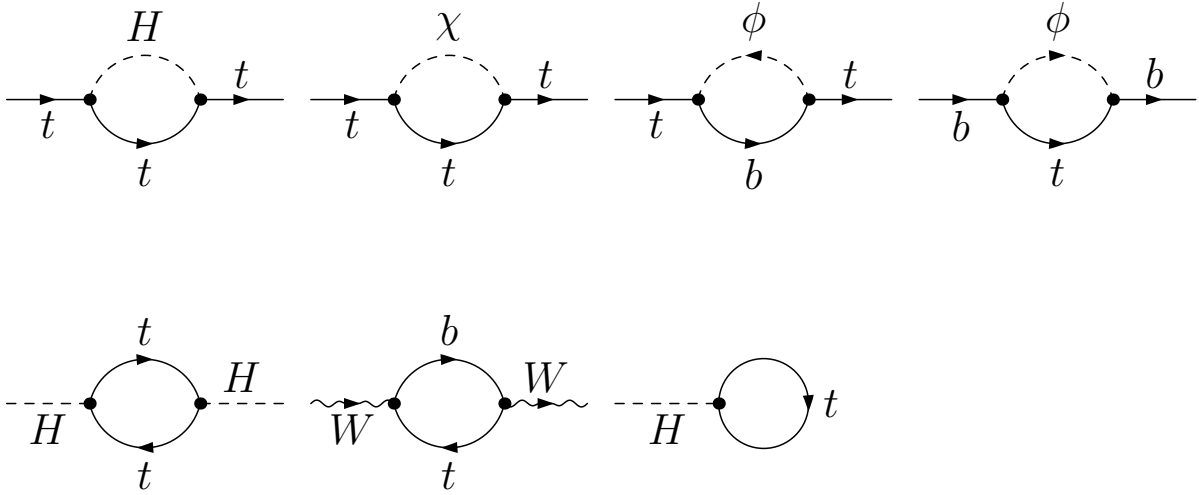}
\end{center}
\caption{\label{Dia1Loop}One-loop self-energy and tadpole diagrams
contributing at order ${\cal O}(G_Fm_t^2)$.}
\end{figure}

In the remainder of this subsection, we list all relevant renormalisation
constants of order ${\cal O}(G_F m_t^2)$.
They are derived by evaluating the diagrams of Fig.~\ref{Dia1Loop} and
applying Eqs.~(\ref{AusdrDmHq1loop}), (\ref{AusdrDZH1loop}),
(\ref{DeltaMF1Loop}), (\ref{AusdrDZbl1loop}), (\ref{AusdrDZbr1loop}),
(\ref{AusdrDmwq}), and (\ref{AusdrDt}).
Since we shall compute the correction of order ${\cal O}(G_F^2 m_t^4)$, these
renormalisation constants are needed through order ${\cal O}(\epsilon)$ in the
expansion in $\epsilon$.
The results read
\begin{eqnarray}
\delta t^{(1)} &=&C_{\epsilon,0}x_{t,0}m_{t,0}^2v_0N_c
\left[\frac{4}{\epsilon}+4+(4+2\zeta(2))\epsilon+{\cal O}(\epsilon^2)\right],
\label{RC1}\\
\delta M_H^{2(1)} &=&C_{\epsilon,0}x_{t,0}m_{t,0}^2N_c
\left[-\frac{12}{\epsilon}-4+(-4-6\zeta(2))\epsilon+{\cal O}(\epsilon^2)
\right],
 \label{RC2}\\
\delta Z_H^{(1)} &=&C_{\epsilon,0}x_{t,0}N_c  
\left[-\frac{2}{\epsilon}+\frac{4}{3}-\zeta(2)\epsilon+{\cal O}(\epsilon^2)
\right],
\label{RC7}\\ 
\frac{\delta m_b^{(1)}}{m_b} &=&C_{\epsilon,0}x_{t,0} 
\left[-\frac{3}{2\epsilon}-\frac{5}{4}+\left(-\frac{9}{8}-\frac{3}{4}\zeta(2
)\right)\epsilon+{\cal O}(\epsilon^2)\right],
\label{RC5}\\ 
\delta Z_{b,L}^{(1)} &=&C_{\epsilon,0}x_{t,0}  
\left[-\frac{1}{\epsilon}-\frac{3}{2}+\left(-\frac{7}{4}-\frac{1}{2}\zeta(2
)\right)\epsilon+{\cal O}(\epsilon^2)\right],
\label{RC8}\\ 
\delta Z_{b,R}^{(1)} &=& 0,
\label{RC9}\\
\frac{\delta m_t^{(1)}}{m_t} &=&C_{\epsilon,0}x_{t,0}  
\left[\frac{3}{2\epsilon}+4+\left(9-\frac{5}{4}\zeta(2
)\right)\epsilon+{\cal O}(\epsilon^2)\right],
\label{RC6}\\ 
\delta M_W^{2(1)} &=&C_{\epsilon,0}x_{t,0}M_{W,0}^2N_c  
\left[-\frac{2}{\epsilon}-1+\left(-\frac{1}{2}-\zeta(2)\right)\epsilon
+{\cal O}(\epsilon^2)\right],
\label{RC3}
\end{eqnarray}
where we use the abbreviations
\begin{eqnarray}
C_\epsilon& =& \left( \frac{4 \pi \mu^2}{m_t^2}e^{-\gamma_E}
\right)^\epsilon,
\nonumber\\
x_t &=& \frac{G_F m_t^2}{8\pi^2\sqrt{2}},
\end{eqnarray}
with $\gamma_E$ being Euler's constant.

\boldmath
\subsection{Correction of order ${\cal O}(G_F m_t^2)$}
\label{SecOneLoop}
\unboldmath

At order ${\cal O}(G_F m_t^2)$, only the one diagram depicted in
Fig.~\ref{DiaHbb}(b) contributes.
Using the notation of Eq.~(\ref{AAufspaltung}), we obtain for the expansion in
$\epsilon$ through order ${\cal O}(\epsilon)$:
\begin{eqnarray}
{\cal A}^{(1)}_{A,0} &=&C_{\epsilon,0}x_{t,0}\frac{m_{b,0}}{v_0} 
\left[-\frac{2}{\epsilon}+2+(2-\zeta(2))\epsilon+{\cal O}(\epsilon^2)\right]
\nonumber\\
{\cal A}^{(1)}_{B,0} &=&C_{\epsilon,0}x_{t,0}\frac{1}{v_0}
\left(-1-\frac{3}{2}\epsilon+{\cal O}(\epsilon^2)\right).
\end{eqnarray}

Expanding Eq.~(\ref{TAusAundZ}) and replacing the bare masses by the
renormalised ones plus their counterterms in Eq.~(\ref{BornUnren}), we
find the transition matrix element to be
\begin{equation} 
{\cal T}^{(1)} = {\cal A}_{A,0}^{(1)} + m_b{\cal A}_{B,0}^{(1)}
+{\cal A}_0^{(0)}
\left(\delta_u^{(1)}+\frac{\delta m_b^{(1)}}{m_b}
+\frac{1}{2} \delta Z_{b,L}^{(1)}+\frac{1}{2} \delta Z_{b,R}^{(1)}\right),
\label{T1Gf}
\end{equation}
where ${\cal A}^{(0)}$ is the amputated matrix element of
Eq.~(\ref{BornUnren})
and
\begin{equation}
\delta_u^{(1)} = \frac{1}{2} \delta Z_H^{(1)}
-\frac{1}{2}\,\frac{\delta M_W^{2(1)}}{M_W^2}
\end{equation}
is the one-loop contribution to the universal counterterm $\delta_u$, which
exhausts the full ${\cal O}(G_F m_t^2)$ corrections for Higgs-boson decays to
fermion-antifermion pairs, except for those into $t\overline{t}$ and
$b\overline{b}$ pairs.
For simplicity, we omitted the spinors on the right-hand side of
Eq.~(\ref{T1Gf}); we shall also do this in the following.
$\delta_u^{(1)}$ and ${\cal T}^{(1)}$ are ultraviolet finite and read
\begin{eqnarray}
\delta_u^{(1)}&=&x_tN_c\frac{7}{6}
\nonumber\\
&=&x_t\frac{7}{2},
\\
{\cal T}^{(1)} &=& {\cal T}^{(0)} x_t \left(-3+N_c\frac{7}{6} \right).
\label{ErgT1}
\end{eqnarray}
The ${\cal O}(G_F m_t^2)$ correction to the decay rate thus becomes
\begin{eqnarray}
\frac{\Gamma^{(1)}}{\Gamma^{(0)}}&=&
x_t\left(-6+N_c\frac{7}{3}\right)
\nonumber\\
&=&x_t,
\end{eqnarray}
where $\Gamma^{(0)}$ is given in Eq.~(\ref{BornTrans}).
The results of this subsection are in accordance with Ref.~\cite{Kniehl:1991}.

\boldmath
\subsection{Correction of order ${\cal O}(G_F^2 m_t^4)$}
\label{CapEW2LoopKorr}
\unboldmath

Expanding Eq.~(\ref{TAusAundZ}) up to the two-loop order and replacing all
bare masses in the tree-level and one-loop amputated matrix elements by the
renormalised masses plus the corresponding counterterms, we find the following
master formula for the transition matrix element
\begin{eqnarray}
{\cal T}^{(2)} &=&
{\cal A}_{A,0}^{(2)} + m_b{\cal A}_{B,0}^{(2)} + {\cal A}_{A,0}^{(1)}
\left(\frac{\delta m_b^{(1)}}{m_b}+\frac{1}{2} \delta Z_{b,L}^{(1)}
+\frac{1}{2} \delta Z_{b,R}^{(1)}\right) 
+m_b{\cal A}_{B,0}^{(1)}\delta Z_{b,L}^{(1)}
\nonumber\\
&&{}+ \left( {\cal A}_{A,0}^{(1)} + m_b {\cal A}_{B,0}^{(1)} \right) \left[
\delta_u^{(1)}
+ 2(1-\epsilon) \frac{\delta m_t^{(1)}} {m_t} 
-\frac{\delta M_W^{2(1)}} {M_W^2} \right]
\nonumber\\
&&{}+ {\cal A}_0^{(0)} \left[ \delta_u^{(2)} + \frac{\delta m_b^{(2)}}{m_b}
+ \frac{1}{2} \delta Z_{b,L}^{(2)} + \frac{1}{2} \delta Z_{b,R}^{(2)} 
+ \delta_u^{(1)}\left(
\frac{\delta m_b^{(1)}}{m_b}
+ \frac{1}{2} \delta Z_{b,L}^{(1)} + \frac{1}{2} \delta Z_{b,R}^{(1)} \right)
\right.\nonumber\\
&&{}+\left.\frac{1}{2}\,\frac{\delta m_b^{(1)}}{m_b}
\left(\delta Z_{b,L}^{(1)} + \delta Z_{b,R}^{(1)}\right)
- \frac{1}{8} \left(\delta Z_{b,L}^{(1)}-\delta Z_{b,R}^{(1)}\right)^2
\right],
\label{AusdrT2Loop}
\end{eqnarray}
where
\begin{equation}
\delta_u^{(2)} = \frac{1}{2} \delta Z_H^{(2)}
-\frac{1}{2}\, \frac{\delta M_W^{2(2)}}{M_W^2} 
- \frac{1}{8} \left(\delta Z_H^{(1)}\right)^2
- \frac{1}{4} \delta Z_H^{(1)} \frac{\delta M_W^{2(1)}}{M_W^2} 
+ \frac{3}{8} \left(\frac{\delta M_W^{2(1)}}{M_W^2} \right)^2 
\label{UnivCT}
\end{equation}
is the universal counterterm.

\subsubsection{Universal counterterm}

\begin{figure}
\begin{center}
\includegraphics[width=\textwidth]{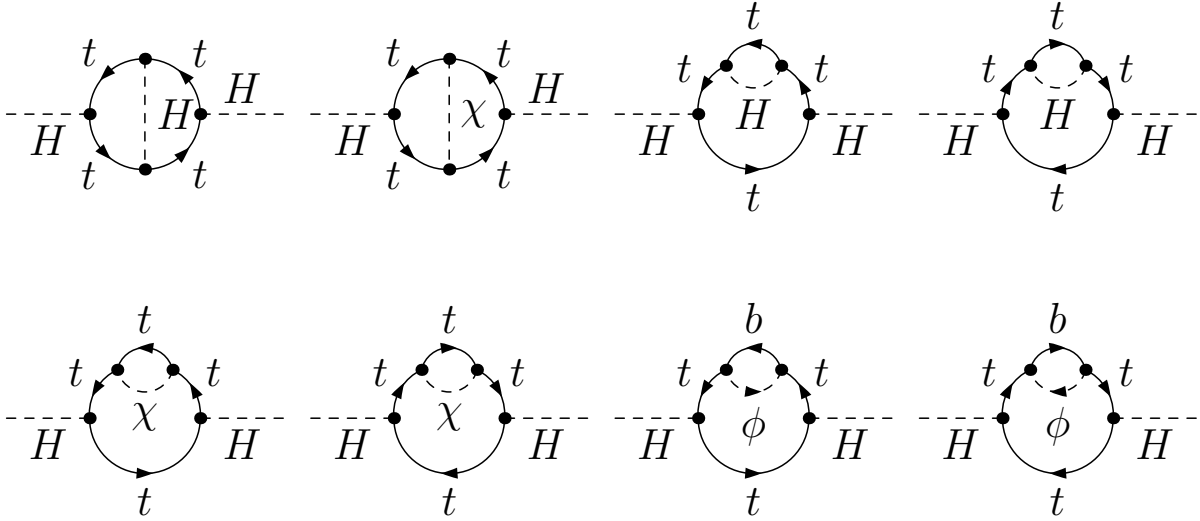}
\end{center}
\caption{\label{DiaH2l}Higgs-boson self-energy diagrams contributing at order
${\cal O}(G_F^2 m_t^4)$.}
\end{figure}

\begin{figure}
\begin{center}
\includegraphics[width=\textwidth]{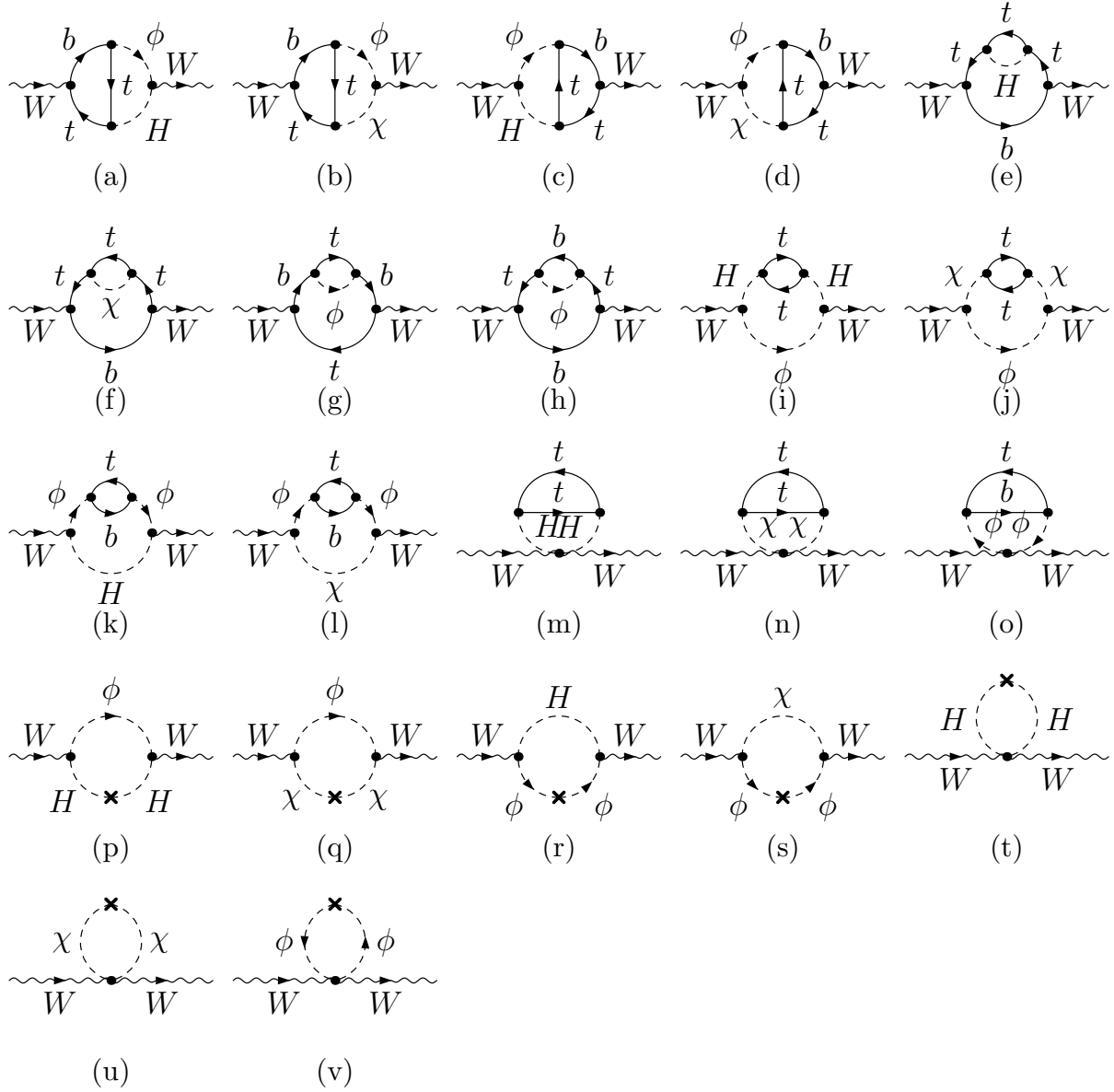}
\end{center}
\caption{\label{DiaW2l}$W$-boson self-energy diagrams contributing at order
${\cal O}(G_F^2 m_t^4)$.
Insertions of $-i\delta M_H^2$ in Higgs-boson lines and of $i\delta t/v_0$ in
$\phi$- or $\chi$-boson lines are indicated by crosses.}
\end{figure}

Let us first calculate the universal counterterm.
To this end, we need the two-loop expressions for $\delta Z_H$ and
$\delta M_W^2$.
The unrenormalised expressions are obtained by evaluating the diagrams in
Figs.~\ref{DiaH2l} and \ref{DiaW2l} and applying Eqs.~(\ref{AusdrDZH2loop})
and (\ref{AusdrDmwq}), the results being
\begin{eqnarray} 
\delta Z_{H,0}^{(2)}&=&C_{\epsilon,0}^2x_{t,0}^2N_c 
\left[\frac{3}{\epsilon^2}-\frac{11}{2\epsilon}-\frac{17}{12} +
5\zeta(2)
+N_c\left(\frac{4}{\epsilon^2}-\frac{16}{3\epsilon}+\frac{16}{9} +
4\zeta(2)\right)+{\cal O}(\epsilon)\right],
\nonumber\\ 
\delta M_{W,0}^{2(2)} &=&C_{\epsilon,0}^2x_{t,0}^2M_{W,0}^2N_c 
\left(\frac{3}{\epsilon^2}+\frac{3}{2\epsilon}
-\frac{69}{4}+17\zeta(2)+{\cal O}(\epsilon)\right),
\end{eqnarray}
in accordance with Ref.~\cite{Djouadi}.
In addition, there are contributions from the renormalisations of the bare
parameters in Eqs.~(\ref{RC7}) and (\ref{RC3}), so that
\begin{eqnarray}
\delta Z_H^{(2)}&=& \delta Z_{H,0}^{(2)} +
\delta Z_H^{(1)} \left[ 2(1-\epsilon)\frac{\delta m_t^{(1)}}{m_t} 
- \frac{\delta M_W^{2(1)}}{M_W^2} \right],
\nonumber\\
\delta M_W^{2(2)}&=& \delta M_{W,0}^{2(2)}+
2(1-\epsilon)\frac{\delta m_t^{(1)}}{m_t}\delta M_W^{2(1)}.
\end{eqnarray}
We are now in a position to specify the universal counterterm at order
${\cal O}(G_F^2 m_t^4)$ as defined in Eq.~(\ref{UnivCT}).
The result is
\begin{eqnarray}
\delta_u^{(2)}&=&x_t^2N_c\left(\frac{29}{2}-6\zeta(2)
+N_c\frac{49}{24}\right)
\nonumber\\
&=&x_t^2\left(\frac{495}{8}-3\pi^2\right).
\label{ErgUnivCT}
\end{eqnarray}

If we convert Eq.~(\ref{ErgUnivCT}) to a mixed renormalisation scheme which
uses on-shell definitions for the particle masses and the definitions of the
modified minimal-subtraction ($\overline{\mathrm{MS}}$) scheme for all other
basic parameters, then we find agreement with Eq.~(15) for $x=0$ in the
paper by Djouadi et al.\ \cite{Djouadi}.
However, the corresponding result for the pure on-shell scheme presented in
their Eq.~(27) for $x=0$ disagrees with our Eq.~(\ref{ErgUnivCT}).
We can trace this discrepancy to the absence in their Eq.~(25) of the
additional finite term $\hat{\delta}_{u}^{(1)}\Delta\rho^{(1)}$ which arises
from the renormalisation of the one-loop result in their Eq.~(7) according to
the prescription in their Eq.~(18).

\subsubsection{Complete transition matrix element}

\begin{figure}
\begin{center}
\includegraphics[width=\textwidth]{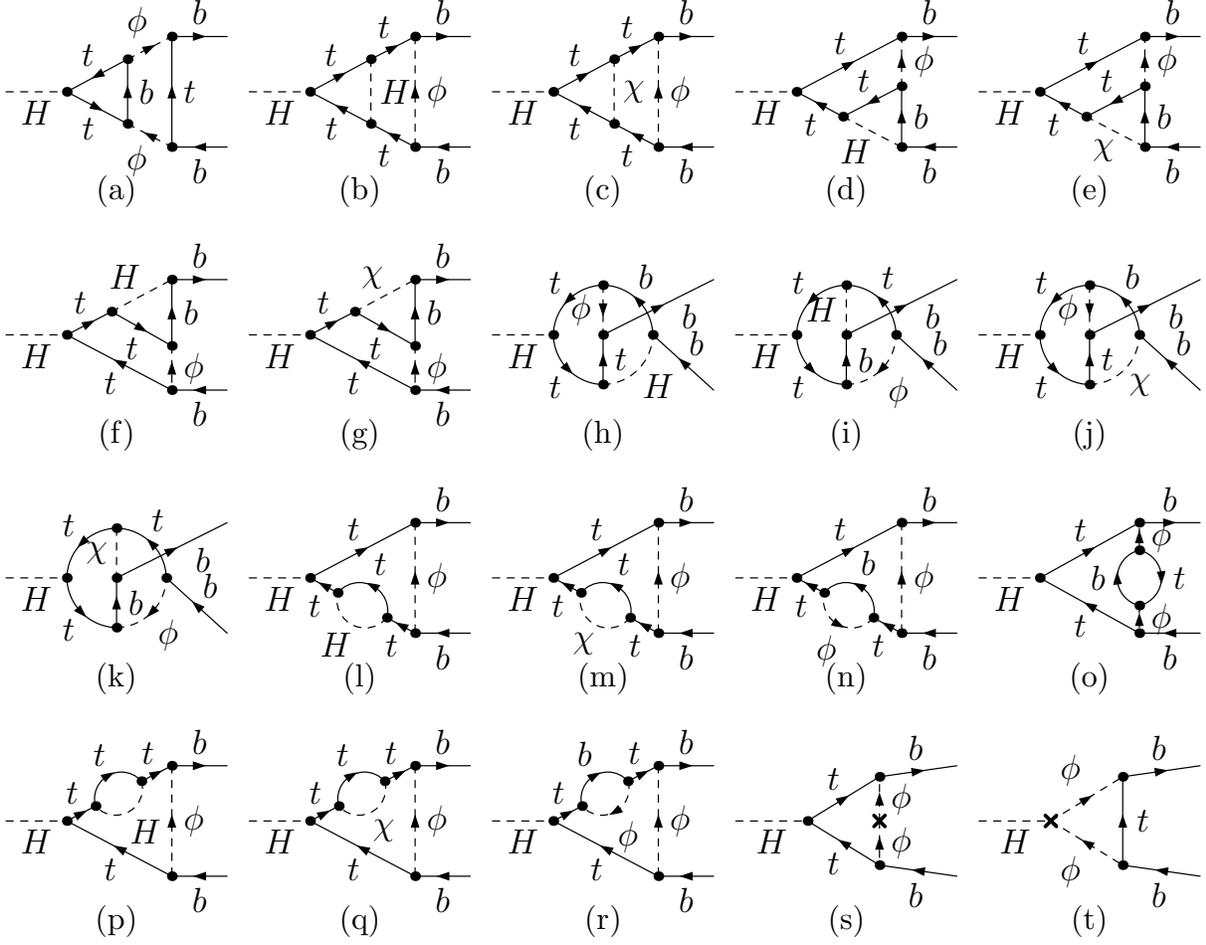}
\end{center}
\caption{\label{DiaHbb2loop}Diagrams contributing to $H\to b\overline{b}$ at
order ${\cal O}(G_F^2 m_t^4)$.
Insertions of $i\delta t/v_0$ in $\phi$-boson lines and of
$-i\left(\delta t/v_0+\delta M_H^2\right)/v_0$ in $H\phi\phi$ vertices are
indicated by crosses.}
\end{figure}

Having provided $\delta_u^{(2)}$, we now turn to the residual
ingredients entering the transition matrix element of Eq.~(\ref{AusdrT2Loop}).
Evaluating the $Hb\overline{b}$ diagrams shown in Fig.~\ref{DiaHbb2loop}, we
find the form factors in Eq.~(\ref{AAufspaltung}) at order
${\cal O}(G_F^2 m_t^4)$ to be
\begin{eqnarray}
{\cal A}_{A,0}^{(2)} &=& \frac{m_{b,0}}{v_0}x_{t,0}^2C_{\epsilon,0}^2
\left[\frac{1}{\epsilon^2}-\frac{5}{\epsilon}-5 +7\zeta(2)
+N_c\left(\frac{2}{\epsilon^2}-\frac{2}{\epsilon}-14 -2\zeta(2)\right)
+{\cal O}(\epsilon)\right],
\nonumber\\
{\cal A}_{B,0}^{(2)} &=& \frac{1}{v_0}x_{t,0}^2C_{\epsilon,0}^2
\left[\frac{2}{\epsilon}+1
+N_c\left(\frac{2}{\epsilon}+9\right)
+{\cal O}(\epsilon)\right].
\end{eqnarray}

\begin{figure}
\begin{center}
\includegraphics[width=\textwidth]{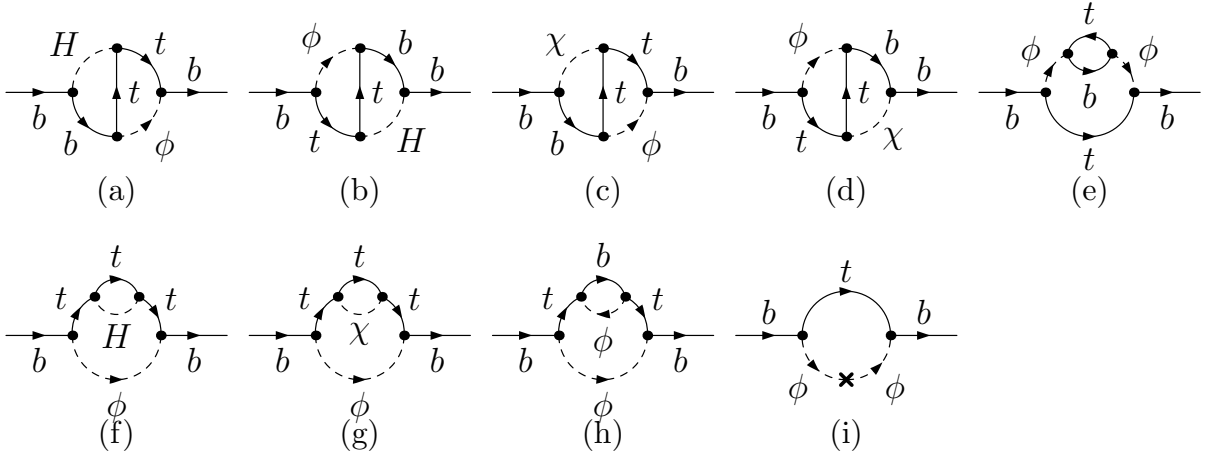}
\end{center}
\caption{\label{DiaB2l}$b$-quark self-energy diagrams contributing at order
${\cal O}(G_F^2 m_t^4)$.
Insertions of $i\delta t/v_0$ in $\phi$-boson lines are indicated by crosses.}
\end{figure}

Evaluating the diagrams depicted in Fig.~\ref{DiaB2l} and using
Eqs.~(\ref{DeltaMF2Loop}), (\ref{AusdrDZbl2loop}), and (\ref{AusdrDZbr2loop}),
we obtain the bottom-quark mass and wave-function renormalisation constants at
order ${\cal O}(G_F^2 m_t^4)$.
The renormalisation constants in terms of bare parameters read
\begin{eqnarray}
\frac{\delta m_{b,0}^{(2)}}{m_b}&=&C_{\epsilon,0}^2x_{t,0}^2
\left[\frac{27}{8\epsilon^2}+\frac{31}{8\epsilon}+\frac{13}{32}+\frac{59}{8}
\zeta(2)
+N_c\left(\frac{3}{2\epsilon^2}+\frac{15}{4\epsilon} +\frac{55}{8}
-\frac{3}{2}\zeta(2)\right)
+{\cal O}(\epsilon)\right],
\nonumber\\
\delta Z_{b,L,0}^{(2)} &=&C_{\epsilon,0}^2x_{t,0}^2
\left[\frac{2}{\epsilon^2}+\frac{7}{2\epsilon}+1+6\zeta(2)
+N_c\left(\frac{1}{\epsilon^2}+\frac{9}{2\epsilon}+\frac{25}{4}
-\zeta(2)\right)
+{\cal O}(\epsilon)\right],
\nonumber\\
\delta Z_{b,R,0}^{(2)} &=&0.
\end{eqnarray}
Additional contributions arise from the replacement of the bare $t$-quark and
$W$-boson masses in Eqs.~(\ref{RC5}), (\ref{RC8}), and (\ref{RC9}), so that
\begin{eqnarray}
\frac{\delta m_b^{(2)}}{m_b} &=&
\frac{\delta m_{b,0}^{(2)}}{m_b}+ \frac{\delta m_b^{(1)}}{m_b}
\left[ 2(1-\epsilon) \frac{\delta m_t^{(1)}}{m_t} 
- \frac{\delta M_W^{2(1)}}{M_W^2} \right],
\nonumber\\
\delta Z_{b,L}^{(2)}&=& \delta Z_{b,L,0}^{(2)} +
\delta Z_{b,L}^{(1)} \left[2(1-\epsilon) \frac{\delta m_t^{(1)}}{m_t} 
- \frac{\delta M_W^{2(1)}}{M_W^2} \right],
\nonumber\\
\delta Z_{b,R}^{(2)}&=&0.
\end{eqnarray}

Now all ingredients for the evaluation of the renormalised transition matrix
element of order ${\cal O}(G_F^2 m_t^4)$ according to Eq.~(\ref{AusdrT2Loop})
are available.
We find
\begin{equation}
{\cal T}^{(2)} = {\cal T}^{(0)} x_t^2\left[
-\frac{29}{2} + N_c(18-6\zeta(2)) + N_c^2\frac{49}{24}\right].
\label{ErgT2}
\end{equation}

Adding Eqs.~(\ref{ErgT0}), (\ref{ErgT1}), and (\ref{ErgT2}), squaring, and
extracting the ${\cal O}(G_F^2 m_t^4)$ term, we have
\begin{eqnarray}
\frac{\Gamma^{(2)}}{\Gamma^{(0)}}&=&
x_t^2\left[-20+N_c(29-12\zeta(2))+N_c^2\frac{49}{9}\right]
\nonumber\\
&=&x_t^2(116-6\pi^2).
\end{eqnarray}

\boldmath
\subsection{Correction of order ${\cal O}(\alpha_s G_F m_t^2)$}
\label{sec:mixed}
\unboldmath

As a by-product of our analysis, we can also compute the
${\cal O}(\alpha_s G_F m_t^2)$ correction to the $H\to b\overline{b}$ decay
width.
The comparison of our result with the literature
\cite{Kniehl:1994ph,KniehlSpira,Kwiatkowski:1994cu} provides a partial check
of our ${\cal O}(G_F^2 m_t^4)$ results.
Note, however, that the calculation considerably simplifies as one passes from
order ${\cal O}(G_F^2 m_t^4)$ to order ${\cal O}(\alpha_s G_F m_t^2)$.
Using our tools, we indeed recover the well-known
${\cal O}(\alpha_s G_F m_t^2)$ results for the universal correction
\cite{Kniehl:1994ph} and the correction to the $H\to b\overline{b}$ decay
width \cite{KniehlSpira,Kwiatkowski:1994cu},
\begin{eqnarray}
\delta_u^{(X_t\alpha_s)}&=&
X_t\frac{\alpha_s}{\pi}C_F N_c\left(-\frac{3}{4}-\frac{\zeta(2)}{2}\right),
\nonumber\\
\frac{\Gamma^{(X_t\alpha_s)}}{\Gamma^{(0)}}&=&
X_t\frac{\alpha_s}{\pi}C_F\left[-12+9\ln \frac{M_H^2}{M_b^2} 
+N_c\left(\frac{15}{4}-\zeta(2)-\frac{7}{2}\ln\frac{M_H^2}{M_b^2}\right)
\right],
\label{asxtonshell}
\end{eqnarray}
respectively, where $X_t=G_FM_t^2/\left(8\pi^2\sqrt{2}\right)$ and
$C_F=(N_c^2-1)/(2N_c)$.
In Eq.~(\ref{asxtonshell}), the bottom- and top-quark masses are denoted with
capital letters, $M_b$ and $M_t$, respectively, to indicate that they are pure
on-shell masses, i.e.\ they are defined in the on-shell scheme also with
regard to quantum chromodynamics (QCD).
The obvious disadvantage of this choice is the appearance of large logarithms
of the type $\ln\left(M_H^2/m_b^2\right)$ starting already in order
${\cal O}(\alpha_s)$, which spoil the convergence behaviour of the
perturbation expansion.
As is well known \cite{Braaten:1980yq}, these logarithms can be resummed into
the running bottom-quark mass, if $m_b$ appearing in Eq.~(\ref{BornTrans}) is
QCD-renormalised in the $\overline{\mathrm{MS}}$ scheme at scale $\mu=M_H$,
by substituting $m_b=\overline{m}_b(M_H)$.
For consistency with the ${\cal O}(G_Fm_t^2)$ and ${\cal O}(G_F^2m_t^4)$
results presented above, which all refer to the electroweak on-shell scheme,
we continue our discussion in a mixed renormalisation scheme where the
on-shell definition of bottom-quark mass is adopted for electroweak
corrections and the $\overline{\mathrm{MS}}$ one for QCD corrections.
Since we wish to treat the masses of the top and bottom quarks on the same
footing, we adopt this mixed scheme for the top-quark mass as well.
Furthermore, the analysis at order ${\cal O}(\alpha_s^2G_Fm_t^2)$
\cite{delu,Chetyrkin:1996ke} reveals that Eq.~(\ref{asxtonshell}) is further
improved according to the renormalisation group if $m_t$ and $\alpha_s$
are taken to be $m_t=\overline{m}_t(m_t)$ and $\alpha_s=\alpha_s^{(n_f)}(m_t)$
with $n_f=6$ quark flavours, respectively.
In this improved renormalisation scheme, Eq.~(\ref{asxtonshell}) takes the
form
\begin{eqnarray}
\delta_u^{(x_t\alpha_s)}&=&x_t\frac{\alpha_s}{\pi}C_FN_c
\left(\frac{19}{12}-\frac{\zeta(2)}{2}\right)
\nonumber\\
&=&x_t\frac{\alpha_s}{\pi}\left(\frac{19}{3}-\frac{\pi^2}{3}\right),
\nonumber\\
\frac{\Gamma^{(x_t\alpha_s)}}{\Gamma^{(0)}}&=&
x_t\frac{\alpha_s}{\pi}C_F
\left[-36+N_c\left(\frac{157}{12}-\zeta(2)\right)\right]
\nonumber\\
&=&x_t\frac{\alpha_s}{\pi}\left(\frac{13}{3}-\frac{2}{3}\pi^2\right).
\label{mixed}
\end{eqnarray}
To the order considered here, we have
\begin{equation}
m_t=M_t\left(1-\frac{\alpha_s^{(6)}(M_t)}{\pi}C_F\right).
\end{equation}

\section{Low-energy theorem}
\label{CapNieder}

In this section, we present an alternative way of calculating all but one of
the $Hb\overline{b}$ diagrams at order ${\cal O}(G_F^2 m_t^4)$ which is based
on the Higgs-boson low-energy theorem \cite{Kniehl:1995tn}.
In fact, the $Hb\overline{b}$ diagrams of Fig.~\ref{DiaHbb2loop}, with the
exception of diagram (t), can be generated from the bottom-quark self-energy
diagrams of Fig.~\ref{DiaB2l} by in turn attaching an external Higgs-boson
line to each of the top-quark lines.
Diagrammatically, this can be represented as follows:
\begin{eqnarray}
\begin{minipage}{56pt}
\begin{picture}(56,40)
  \ArrowLine(0,10)(56,10)
  \Text(28,14)[b]{$t(q)$}
\end{picture}
\end{minipage}
\quad&\longrightarrow&\quad
\begin{minipage}{112pt}
\begin{picture}(112,40)
  \ArrowLine(0,10)(56,10)
  \Text(28,14)[b]{$t(q)$}
  \ArrowLine(56,10)(112,10)
  \Text(84,14)[b]{$t(q)$}
  \DashLine(56,10)(56,40){5}
  \Text(58,40)[lt]{$H$}
\end{picture}
\end{minipage}
\nonumber \\
\frac{i}{\slashed{q}-m_{t,0}} \quad&\longrightarrow&\quad 
\frac{i}{\slashed{q}-m_{t,0}}
\,\frac{-im_{t,0}}{v_0}  \,\frac{i}{\slashed{q}-m_{t,0}}.
\end{eqnarray}
Here, we also made use of the fact that, in the large-$m_t$ approximation,
the external Higgs boson does not carry any four-momentum into the respective
diagram.
Thanks to the identity
\begin{equation}
\frac{i}{\slashed{q}-m_{t,0}}\,
 \frac{-im_{t,0}}{v_0}\, \frac{i}{\slashed{q}-m_{t,0}} =
\frac{m_{t,0}}{v_0}\,\frac{\partial}{\partial m_{t,0}} \left( \frac{i}{\slashed{q}-m_{t,0}}
\right),
\end{equation}
the amputated matrix element of $H\to b\overline{b}$ is in the large-$m_t$
limit related to the bottom-quark self-energy as
\begin{equation} \label{FormelNieder}
{\cal A}_0 = \frac{m_{t,0}}{v_0} 
\,\frac{\partial}{\partial m_{t,0}} \Sigma_b,
\end{equation}
where it is understood that the differential operator only acts on masses
which stem from propagators, not to those occurring in vertices, and that all
quantities in Eq.~(\ref{FormelNieder}) are taken to be bare.
Exploiting the structures underlying Eqs.~(\ref{FermSelbstDef}) and
(\ref{AAufspaltung}), Eq.~(\ref{FormelNieder}) can be decomposed into two
scalar equations.
Identifying the four-momentum $q$ in Eq.~(\ref{FermSelbstDef}) with $q_2$ in
Eq.~(\ref{AAufspaltung}) and noticing that $q_2=-q_1$ in the soft-Higgs limit,
we have
\begin{eqnarray}
{\cal A}_{A,0}&=& m_{b,0}\frac{m_{t,0}}{v_0}
\,\frac{\partial}{\partial m_{t,0}}\Sigma_{b,S},
\nonumber\\
{\cal A}_{B,0}&=&\frac{1}{2}\,\frac{m_{t,0}}{v_0}\,
\frac{\partial}{\partial m_{t,0}}\Sigma^{b,L}.
\label{let}
\end{eqnarray}
The fact that the $H\to b\overline{b}$ amplitude does not contain a term
proportional to $(\slashed{q}_2-\slashed{q}_1)\omega_+$ is reflected by the
fact that the right-handed part of the bottom-quark self-energy,
$\Sigma_{b,R}$, vanishes to the orders considered in this paper.

The results for ${\cal A}_{A,0}$ and ${\cal A}_{B,0}$ obtained through
Eq.~(\ref{let}) indeed agree with the direct evaluation of the respective
diagrams in Fig.~\ref{DiaHbb2loop}.

\section{Numerical results}
\label{Numerics}

Finally, we explore the phenomenological implications of our results.
Adopting from Ref.~\cite{PDG} the values $G_F=1.16637\times10^{-5}$~GeV$^{-2}$,
$\alpha_s^{(5)}(M_Z)=0.1176$, $M_Z=91.1876$~GeV, and $M_t=174.2$~GeV for our
input parameters, so that $\alpha_s^{(6)}(m_t)=0.1076$ and $m_t=166.2$~GeV,
we evaluate the relative corrections $\Gamma^{(x)}/\Gamma^{(0)}$ to the
$H\to b\overline{b}$ decay width to orders $x=G_F m_t^2$, $G_F^2m_t^4$, and
$\alpha_sG_Fm_t^2$.
For comparison, we also evaluate the relative corrections to the $H\to l^+l^-$
and $H\to q\overline{q}$ decay widths, where $l=e,\mu,\tau$ and $q=u,d,s,c$,
which, to the orders considered here, are given by
\begin{eqnarray}
\Delta_l&=&(1+\delta_u)^2-1
\nonumber\\
&=&2\delta_u^{(1)}
+2\delta_u^{(2)}+\left(\delta_u^{(1)}\right)^2
+2\delta_u^{(x_t\alpha_s)},
\nonumber\\
\Delta_q&=&(1+\Delta_\mathrm{QCD})(1+\delta_u)^2-1
\nonumber\\
&=&\Delta_\mathrm{QCD}+2\delta_u^{(1)}
+2\delta_u^{(2)}+\left(\delta_u^{(1)}\right)^2
+2\delta_u^{(x_t\alpha_s)}
+2\Delta_\mathrm{QCD}\delta_u^{(1)},
\end{eqnarray}
where \cite{Braaten:1980yq}
\begin{equation}
\Delta_\mathrm{QCD} = \frac{\alpha_s}{\pi}C_F \frac{17}{4}
\end{equation}
is the ${\cal O}(\alpha_s)$ correction in the limit $m_q\ll M_H$, with
$m_q=\overline{m}_q(M_H)$.

The results are listed in Table~\ref{tab:num}.
We observe that the ${\cal O}(G_F^2m_t^4)$ correction to $\Gamma^{(0)}$
increases the enhancement due to the ${\cal O}(G_F m_t^2)$ one by about 16\%
and has more than twice the magnitude of the negative
${\cal O}(\alpha_sG_Fm_t^2)$ one.
\begin{table}[t]
\begin{center}
\caption{\label{tab:num}Numerical values of the relative corrections
$\Delta_l^{(x)}$, $\Delta_q^{(x)}$, and $\Gamma^{(x)}/\Gamma^{(0)}$ to the
$H\to l^+l^-$, $H\to q\overline{q}$, and $H\to b\overline{b}$ decay widths,
respectively, at orders $x=G_Fm_t^2$, $G_F^2m_t^4$, and $\alpha_sG_Fm_t^2$.}
\begin{tabular}{|c|ccc|}
\hline
Order $x$ & $\Delta_l^{(x)}$ & $\Delta_q^{(x)}$ & $\Gamma^{(x)}/\Gamma^{(0)}$
\\
\hline
$\mathcal{O}(G_Fm_t^2)$ & $+2.021\%$ & $+2.021\%$ & $+0.289\%$ \\
$\mathcal{O}(G_F^2m_t^4)$ & $+0.064\%$ & $+0.064\%$ & $+0.047\%$ \\
$\mathcal{O}(\alpha_sG_Fm_t^2)$ & $+0.060\%$ & $+0.452\%$ & $-0.022\%$ \\
\hline
\end{tabular}
\end{center}
\end{table}

\section{Conclusions}
\label{CapZusammenfassung}

We analytically calculated the dominant electroweak two-loop correction, of
order\break ${\cal O}(G_F^2m_t^4)$, to the $H\to b\overline{b}$ decay width of
an intermediate-mass Higgs boson, with $M_H\ll m_t$.

We performed various checks for our analysis.
The ultraviolet divergences cancelled through genuine two-loop renormalisation.
Our final result is devoid of infrared divergences related to infinitesimal
scalar-boson masses.
We reproduced those $Hb\overline{b}$ triangle diagrams where the external
Higgs boson is coupled to an internal top-quark line, which we had computed
directly, through application of a low-energy theorem.
After switching to a hybrid renormalisation scheme, our ${\cal O}(G_F^2m_t^4)$
result for the universal correction $\delta_u$ agrees with
Ref.~\cite{Djouadi}.
Using our techniques, we also recovered the ${\cal O}(\alpha_sG_Fm_t^2)$
correction to the $H\to b\overline{b}$ decay width as well as the universal
correction $\delta_u$ in this order.

The ${\cal O}(G_F^2m_t^4)$ correction to the $H\to b\overline{b}$ decay width 
amplifies the familiar enhancement due to the ${\cal O}(G_Fm_t^2)$ correction
by about $+16\%$ and thus more than compensates the screening by about $-8\%$
through QCD effects of order ${\cal O}(\alpha_sG_Fm_t^2)$.

\section*{Acknowledgements}

We like to thank Paolo Gambino, Jan Piclum, Florian Schwennsen, and Matthias
Steinhauser for fruitful discussions.
This work was supported in part by the German Federal Ministry for Education
and Research BMBF through Grant No.\ 05~HT6GUA and by the German Research
Foundation DFG through Graduate School No.\ GRK~602 {\it Future
Developments in Particle Physics}.


\begin{thebibliography}{10}

\bibitem{Barate:2003sz}
ALEPH Collaboration, DELPHI Collaboration, L3 Collaboration,
OPAL Collaboration and The LEP Working Group for Higgs Boson Searches,
R.~Barate, et al.,
Phys.\ Lett.\ B 565 (2003) 61.

\bibitem{LEPEWWG}
LEP Electroweak Working Group, D.~Abbaneo, et al.,
Report No.\ LEPEWWG/2005-01; see also
URL: {\tt http://lepewwg.web.cern.ch/LEPEWWG/}.

\bibitem{Kniehl:2001jy}
B.A.~Kniehl,
Int.\ J.\ Mod.\ Phys.\ A 17 (2002) 1457.

\bibitem{Kniehl:1993ay}
B.A.~Kniehl,
Phys.\ Rept.\  240 (1994) 211;\\
M.~Spira,
Fortsch.\ Phys.\ 46 (1998) 203.

\bibitem{Braaten:1980yq}
E.~Braaten, J.P.~Leveille,
Phys.\ Rev.\ D 22 (1980) 715;\\
N.~Sakai,
Phys.\ Rev.\ D 22 (1980) 2220;\\
T.~Inami, T.~Kubota,
Nucl.\ Phys.\ B 179 (1981) 171;\\
M.~Drees, K.~Hikasa,
Phys.\ Lett.\ B 240 (1990) 455;\\
M.~Drees, K.~Hikasa,
Phys.\ Lett.\ B 262 (1991) 497, Erratum.

\bibitem{Kniehl:1991}
J.~Fleischer, F.~Jegerlehner,
Phys.\ Rev.\ D 23 (1981) 2001;\\
D.Yu.~Bardin, B.M.~Vilenski\u\i, P.Kh.~ Khristova,
Yad.\ Fiz.\ 53 (1991) 240 [Sov.\ J. Nucl.\ Phys.\ 53 (1991) 152];\\
B.A.~Kniehl,
Nucl.\ Phys.\ B 376 (1992) 3;\\
A.~Dabelstein, W.~Hollik,
Z. Phys.\ C 53 (1992) 507.

\bibitem{Gorishnii:1991zr}
S.G.~Gorishny, A.L.~Kataev, S.A.~Larin, L.R.~Surguladze,
Mod.\ Phys.\ Lett.\ A 5 (1990) 2703;\\
S.G.~Gorishny, A.L.~Kataev, S.A.~Larin, L.R.~Surguladze,
Phys.\ Rev.\ D 43 (1991) 1633;\\
A.L.~Kataev, V.T.~Kim,
Mod.\ Phys.\ Lett.\ A 9 (1994) 1309.

\bibitem{Surguladze:1994gc}
L.R.~Surguladze,
Phys.\ Lett.\ B 341 (1994) 60.

\bibitem{Kniehl:1994vq}
B.A.~Kniehl,
Phys.\ Lett.\ B 343 (1995) 299.

\bibitem{Chetyrkin:1995pd}
K.G.~Chetyrkin, A.~Kwiatkowski,
Nucl.\ Phys.\ B 461 (1996) 3.

\bibitem{Chetyrkin:1996sr}
K.G.~Chetyrkin,
Phys.\ Lett.\ B 390 (1997) 309.

\bibitem{Chetyrkin:1997vj}
K.G.~Chetyrkin, M.~Steinhauser,
Phys.\ Lett.\ B 408 (1997) 320.

\bibitem{Kniehl:1994ph}
B.A.~Kniehl, A. Sirlin,
Phys.\ Lett.\ B 318 (1993) 367;\\
B.A.~Kniehl,
Phys.\ Rev.\ D 50 (1994) 3314;\\
A.~Djouadi, P.~Gambino,
Phys.\ Rev.\ D 51 (1995) 218.

\bibitem{KniehlSpira}
B.A.~Kniehl, M.~Spira,
Nucl.\ Phys.\ B 432 (1994) 39.

\bibitem{Kwiatkowski:1994cu}
A.~Kwiatkowski, M.~Steinhauser,
Phys.\ Lett.\ B 338 (1994) 66;\\
A.~Kwiatkowski, M.~Steinhauser,
Phys.\ Lett.\ B 342 (1995) 455, Erratum.

\bibitem{delu}
B.A.~Kniehl, M.~Steinhauser,
Nucl.\ Phys.\ B 454 (1995) 485;\\
B.A.~Kniehl, M.~Steinhauser,
Phys.\ Lett.\ B 365 (1996) 297.

\bibitem{Chetyrkin:1996ke}
K.G.~Chetyrkin, B.A.~Kniehl, M.~Steinhauser,
Phys.\ Rev.\ Lett.\ 78 (1997) 594;\\
K.G.~Chetyrkin, B.A.~Kniehl, M.~Steinhauser,
Nucl.\ Phys.\ B 490 (1997) 19.

\bibitem{Djouadi}
A.~Djouadi, P.~Gambino, B.A.~Kniehl,
Nucl.\ Phys.\ B 523 (1998) 17.

\bibitem{prl}
M.~Butensch\"on, F.~Fugel, B.A.~Kniehl,
Phys.\ Rev.\ Lett.\ 98 (2007) 071602.

\bibitem{Hahn:2000kx}
T.~Hahn,
Comput.\ Phys.\ Commun.\ 140 (2001) 418.

\bibitem{MATAD}
M.~Steinhauser,
Comput.\ Phys.\ Commun.\ 134 (2001) 335.

\bibitem{FORM}
J.A.M.~Vermaseren,
Symbolic Manipulation with FORM, Computer Algebra Netherlands, Amsterdam,
1991.

\bibitem{Kniehl:1998fn}
B.A.~Kniehl, A.~Sirlin,
Phys.\ Rev.\ Lett.\ 81 (1998) 1373;\\
B.A.~Kniehl, A.~Sirlin,
Phys.\ Lett.\ B 440 (1998) 136;\\
B.A.~Kniehl, C.P.~Palisoc, A.~Sirlin,
Nucl.\ Phys.\ B 591 (2000) 296;\\
B.A.~Kniehl, A.~Sirlin,
Phys.\ Lett.\ B 530 (2002) 129.

\bibitem{Hollik:1988ii}
W.F.L.~Hollik,
Fortsch.\ Phys.\ 38 (1990) 165.

\bibitem{Faisst}
M.~Faisst,
Diploma thesis, University of Karlsruhe, 2000.

\bibitem{Denner}
A.~Denner,
Fortsch.\ Phys.\ 41 (1993) 307.

\bibitem{Smirnov}
V.A.~Smirnov,
Applied Asymptotic Expansions in Momenta and Masses, Springer, Heidelberg,
Germany, 2001.

\bibitem{Kniehl:1995tn}
A.I.~Va\u\i nshte\u\i n, M.B.~Voloshin, V.I.~Zakharov, M.A.~Shifman,
Yad.\ Fiz.\ 30 (1979) 1368 [Sov.\ J. Nucl.\ Phys.\ 30 (1979) 711];\\
A.I.~Va\u\i nshte\u\i n, V.I.~Zakharov, M.A.~Shifman,
Usp.\ Fiz.\ Nauk 131 (1980) 537 [Sov.\ Phys.\ Usp.\ 23 (1980) 429];\\
M.B.~Voloshin,
Yad.\ Fiz.\ 44 (1986) 738 [Sov.\ J. Nucl.\ Phys.\ 44 (1986) 478];\\
M.A.~Shifman,
Usp.\ Fiz.\ Nauk 157 (1989) 561 [Sov.\ Phys.\ Usp.\ 32 (1989) 289];\\
B.A.~Kniehl, M.~Spira,
Z. Phys.\ C 69 (1995) 77;\\
W.~Kilian,
Z.\ Phys.\ C 69 (1995) 89;\\
M.~Spira, A.~Djouadi, D.~Graudenz, P.M.~Zerwas,
Nucl.\ Phys.\ B 453 (1995) 17.

\bibitem{Consoli:1989fg}
M.~Consoli, W.~Hollik, F.~Jegerlehner,
Phys.\ Lett.\ B 227 (1989) 167.

\bibitem{PDG}
Particle Data Group, W.-M.~Yao, et al.,
J. Phys.\ G 33 (2006) 1.

\end{thebibliography}
\end{document}